\documentclass[12pt]{article}
\usepackage[margin=0.95 in]{geometry}
\usepackage{amsmath} 
\usepackage{amssymb,amsfonts}
\usepackage[all]{xy}
\usepackage{graphicx}
\usepackage[utf8x]{inputenc}
\usepackage{amsmath}
\usepackage{amssymb}
\usepackage{float}
\usepackage{array}
\usepackage{tikz}
\usepackage{mathtools}
\usepackage{mathrsfs} 
\usepackage[hidelinks]{hyperref}
\usepackage{cite}
\numberwithin{equation}{section}
\numberwithin{equation}{section}
\numberwithin{table}{section}
\begin{document}

\thispagestyle{empty}

\vspace*{3cm}
{}

\noindent
{\LARGE \bf 
The Topological Symmetric Orbifold}
\vskip .4cm
\noindent
\linethickness{.06cm}
\line(10,0){447}
\vskip 1.1cm
\noindent
\noindent
{\large \bf Songyuan Li and Jan Troost}
\vskip 0.25cm
{\em 
\noindent
 Laboratoire de Physique de l’\'Ecole Normale Sup\'erieure \\ 
 \hskip -.05cm
 CNRS, ENS, Universit\'e PSL,  Sorbonne Universit\'e, Universit\'e de Paris, 75005 Paris, France
}
\vskip 1.2cm

\vskip0cm

\noindent
{\sc Abstract: } {We analyze  topological orbifold conformal field theories on the symmetric product of a complex surface $M$. By exploiting the mathematics literature we show that a canonical quotient of the operator ring has structure constants given by  Hurwitz numbers. This proves a conjecture  in the physics literature on extremal correlators.
Moreover, it allows to leverage results on the combinatorics of the symmetric group to compute more structure constants explicitly.
We recall that the full orbifold chiral ring is given by a symmetric orbifold Frobenius algebra. This construction  enables the computation of  topological genus zero and genus one correlators, and to prove the vanishing of higher genus contributions. The efficient description of all topological correlators sets the stage for a proof of a topological AdS/CFT correspondence. Indeed, we propose a concrete mathematical incarnation of the proof, relating Gromow-Witten theory in the bulk  to the quantum cohomology of the Hilbert scheme on the boundary.}

\vskip 1cm

\pagebreak

\newpage
\setcounter{tocdepth}{2}
\tableofcontents

\section{Introduction}
The symmetric product orbifold conformal field theory  on 
a two-dimensional complex surface $M$ plays an important role in the anti-de Sitter/conformal field theory correspondence in three bulk dimensions \cite{Maldacena:1997re}.  It is a partly solvable  theory that may lie in the moduli space of the conformal field theory dual to string theory in $AdS_3 \times S^3 \times M$. This incarnation of the holographic correspondence has seen a  successful comparison of  correlation functions at infinite central charge \cite{Dabholkar:2007ey,Gaberdiel:2007vu}. Moreover, at maximal bulk curvature, a broad matching of the spectrum and correlation functions has been achieved \cite{Eberhardt:2019ywk,Eberhardt:2020akk}. 

It is natural to ask whether the agreement of the correlators can be extended to  finite central charge. A number of boundary correlators have been computed at finite central charge \cite{Jevicki:1998bm,Lunin:2000yv,Lunin:2001pw}, but the bulk calculations seem hard to perform. Naively,  one would need to get a reasonable handle on (at least) genus one correlation functions in interacting as well as non-compact conformal field theories.  This task would require significant technical advances.

A related question is whether there is a topologically twisted version of this correspondence where both bulk and boundary correlation functions simplify, and where one may hope that the sum over intermediate states necessary in bulk loop calculations significantly truncates.
In principle, the twist of the boundary conformal field theory is well-understood  
\cite{Witten:1988ze,Eguchi:1990vz,Witten:1988xj}, but a detailed description of the resulting theory is lacking in the physics literature, despite the fact that sharp insights were obtained  \cite{Rastelli:2005ph,Pakman:2009ab}. Twisting the bulk theory is the subject of ongoing research \cite{Sugawara:1999fq,Rastelli:2005ph,Li:2019qzx,Costello:2020jbh,Li:2020nei}.

In this paper, we further explore what is known about the correlators of the topologically twisted symmetric product conformal field theory. To understand the answer to this  question, we reformulate both the results that are present on this problem in the mathematics literature, as well as the results in the physics literature. While the two have been rather convincingly matched where it concerns the spectrum of the topological conformal field theory, they have developed largely independently in as far as  the structure constants of the ring of operators  are concerned. Thus, we wish to  simplify and connect these results such that they may shine light on both the  mathematical as well as the physical side of the problem. 
We believe that the efficient mathematical description of the operator ring of the boundary theory  also provides a scheme for the bulk analysis, and therefore for a proof of a topological subsector of an AdS/CFT correspondence. 

The plan of the paper is as follows.
In section \ref{ComplexPlane}, we  review a large number of results on the cohomology ring of the Hilbert scheme of points on the complex plane $\mathbb{C}^2$. We point out that the ring is a canonical quotient ring of the chiral ring of the symmetric orbifold conformal field theory on (quasi-projective) compact complex surfaces $M$. As such it captures part of the full orbifold chiral ring structure associated to compact surfaces $M$. Indeed, we argue that the quotient ring
codes a subset of extremal correlators and use it to prove their conjectured form. Since the quotient ring structure is captured by permutation group combinatorics, the latter mathematical domain provides  new explicit expressions for symmetric orbifold correlators. In section \ref{M}, we extend our review of the mathematics literature to the cohomology ring of the Hilbert scheme of points on the (quasi-projective) complex surfaces $M$. The ring is described by a (symmetric orbifold) Frobenius algebra that can be constructed on the basis of the cohomology ring of $M$ combined with permutation group combinatorics. This gives a compact description of the chiral ring of the topologically twisted symmetric orbifold conformal field theory $Sym_n(M)$ of the manifold $M$. We compute a few correlators using the mathematical formalism and match them onto the known physical correlators. The large $n$ behaviour as well as the nature of loop corrections in the topological theory are clarified using further theorems, and new classes of correlation functions are computed. We conclude in section \ref{Conclusions} with a summary and comments on how these insights may lead to a proof of a topological AdS/CFT correspondence. Importantly, we propose a mathematical incarnation of the proof that includes non-perturbative corrections, relating Gromow-Witten theory in the bulk to the quantum cohomology of the Hilbert scheme. Appendix \ref{LuninMathurCorrelators} contains a technical bridge to the physics literature.

\section{The Complex Plane}
\label{ComplexPlane}
A standard example of a symmetric product space in the mathematics literature is the symmetric product of the two-dimensional complex plane $M=\mathbb{C}^2$, or more precisely, the Hilbert scheme of points on the complex plane. An underlying reason is that the Hilbert scheme of points is both non-trivial and non-singular for two-dimensional complex surfaces \cite{Grothendieck,Fogarty,Beauville}. A thorough understanding of this space \cite{Briancon,EllingsrudStroemme,NakajimaLectures,Grojnowski,Lehn,LehnSorgerSymmetric,Vasserot} has lead to significant progress both in describing the cohomology of Hilbert schemes of points on generic complex surfaces, as well as their cohomology rings \cite{Goettsche,Chen:2000cy,Ruan:2000mz,LehnSorgerCupProduct,FantechiGoettscheOrbifoldCohomology}. See e.g. \cite{Bertin} for an introduction. On the physics side of this domain, the discussion of the Hilbert scheme of points on the complex plane $\mathbb{C}^2$  is largely absent in the two-dimensional conformal field theory literature (although it is  present in the literature on moduli spaces of instantons) since the more standard set-up is to study conformal field theories with a discrete spectrum, which requires the complex target space  $M$ to be compact. The compact surfaces $K3$ as well as $T^4$ are omnipresent in string theory compactifications, and the corresponding physical conformal field theories are well-studied. These studies have been extended to their symmetric orbifold products. In this first section, we wish to bridge the gap between the standard model in mathematics, namely the Hilbert scheme of points on the complex plane $\mathbb{C}^2$ and the topologically twisted symmetric orbifold conformal field theory on the complex plane $\mathbb{C}^2$. 

Thus, we start out with a mathematical description of the cohomology of the Hilbert scheme of points on the  plane, and the cohomology ring. Next, we define the symmetric orbifold conformal field theory, the operators in the twisted version of the theory, the cohomology, as well as the operator product of two cohomology elements. The conclusion will be that the two rings naturally match.  Along the way, we will have learned a dictionary, as well as various subtleties that arise when discussing non-compact models. Moreover, the isomorphism permits us to obtain a large number of new results on the physical model.

\subsection{The Hilbert Scheme of Points on the Complex Plane}
\label{HilbertSchemeComplexPlane}
We recall the definition of the Hilbert scheme of points on the plane $\mathbb{C}^2$ and its relation to the symmetric orbifold space. We review the calculation of its cohomology as well as the cup product in the cohomology ring.
See \cite{Bertin,NakajimaLectures} for  pedagogical introductions to the Hilbert scheme of points.
\subsubsection{The Hilbert Scheme}

The Hilbert scheme of points $Hilb^n(\mathbb{C}^2)$  on the complex plane  is the space of ideals  of co-dimension $n$ in the space of polynomials $\mathbb{C}[x,y]$ \cite{Grothendieck}. The co-dimension one ideals are  seen to be parameterized by the two-plane itself. The co-dimension two ideals correspond to either two distinct points or coinciding points together with an orientation. In general, the Hilbert scheme of points corresponds to a  cover of the orbifold space $(\mathbb{C}^2)^n/S_n$ where $S_n$ is the group that permutes the $n$ copies of the complex two-plane. The projection is the Hilbert-Chow morphism. An important point is that the cover is regular when one studies the Hilbert scheme of points of a complex surface $M$ \cite{Fogarty}.
The extra orientation data captured by the ideals of the polynomial ring desingularize the orbifold variety, and at the same time, correspond to the twisted sectors of the conformal field theory which is also regular when defined in accord with the axioms of two-dimensional conformal field theory \cite{Dixon:1985jw,Dixon:1986jc,Chen:2000cy,Ruan:2000mz}. It is often handy to treat the Hilbert schemes for all values of $n$ simultaneously. Physically, this is akin to studying a second quantized string theory \cite{Dijkgraaf:1996xw}.

\subsubsection{The Cohomology}
The final result for the cohomology of the Hilbert scheme of points on the complex two-plane is simple \cite{EllingsrudStroemme,NakajimaLectures,Haiman98,Bertin}. One method to compute the cohomology is to introduce a perfect Morse function that projects contributions to the cohomology to the fixed points under a $\mathbb{C}^\ast \times\mathbb{C}^\ast$ two-torus action that multiplies the variables $(x,y)$ by a non-zero complex number each \cite{NakajimaLectures}. The points of the Hilbert scheme that are invariant under the action are the monomial ideals, namely those ideals that are generated by monomials $x^a y^b$. When we divide the polynomial ring by the monomial ideal we find a vector space that has a basis of monomials that we can capture in a Young diagram. For a given value of $n$, we choose a partition $[\lambda]=[1^{m_1} 2^{m_2} \dots] =(\lambda_1 \ge \lambda_2 \ge \dots \ge \lambda_l >0)$ of $n=\sum_{i=1} i m_i$ with length $l(\lambda)=\sum_i m_i$. The corresponding Young diagram is given by putting $\lambda_1$ boxes on the bottom row, $\lambda_2$ boxes on the second row and so on. The corresponding basis of the ring divided by the ideal is given by monomials $x^p y^q$
where $(p,q)$ take values in the set $
M_\lambda = \{ (p, q) \in \mathbb{N}^2: 0 \le q < \lambda_{p+1} \}$.
The contribution of a given fixed point to the cohomology is determined by the character of the torus action on the tangent space to the fixed point \cite{Bertin,NakajimaLectures}. 
After a calculation \cite{NakajimaLectures,Haiman98}, one finds that each fixed point partition $\mu \, \vdash n$ contributes to the Poincar\'e polynomial $P(t)$ as:
\begin{equation}
P(t) ((\mathbb{C}^2)^{[n]})= \sum_{\mu \, \vdash n} t^{2(n-l(\mu))} \, , \label{SumFormula}
\end{equation}
where the power of $t$ keeps track of the degree of the cohomology elements.
When we sum the Poincar\'e polynomials of all Hilbert schemes over $n$ with dummy variable $q$ to the power $n$, then we  
 find the generator of Poincar\'e polynomials \cite{NakajimaLectures}
\begin{eqnarray}
P(t,q) &=& \sum_{n \ge 0} q^n P_t((\mathbb{C})^{[n]}) = \prod_{m=1}^{\infty} \frac{1}{1-t^{2m-2} q^m} \, . 
\label{PoincarePolynomialPlane}
\end{eqnarray}
We have $b_{2i} \left( (\mathbb{C}^2)^{[n]} \right) = p(n,n-i)$
\cite{EllingsrudStroemme}, namely the number of partitions of $n$ into $n-i$ parts.
Indeed, when we expand the denominator in the generating function (\ref{PoincarePolynomialPlane}), we need to match the sum of the powers of $q$ to the number of copies $n$. Thus, we count partitions of $n$. Moreover, the power of $t$ is twice the power of $q$, except that we subtract two for each factor and therefore in total twice the length of the partition as in equation (\ref{SumFormula}).

\subsubsection{The Cup Product}
\label{CupProductComplexPlane}
\label{CountingConjugacyClasses}
While the Betti numbers of the Hilbert scheme have been known for quite some time, the algebraic geometric understanding of the cohomology as a Fock space is more recent \cite{NakajimaLectures,Grojnowski}. This insight was further exploited to describe the ring structure constants of the cohomology ring \cite{Lehn,LehnSorgerSymmetric,Vasserot} in elementary terms. The final result of the algebraic geometric analyses is as follows  \cite{LehnSorgerSymmetric}.

To each partition $\mu$, one associates a conjugacy class of permutations $[\pi]$ which has cycle lengths given by the partition $\mu$. We equip the space of functions on the permutations $S_n$ with a convolution product $\ast$:
\begin{equation}
(f \ast g)(\pi) = \sum_{\sigma \in S_n} f(\pi \sigma^{-1}) g(\sigma) \, , \label{Convolution}
\end{equation}
that is inherited by the space of functions on the conjugacy classes. 
Moreover, each permutation $\pi$ is assigned a degree $|\pi|$ which is equal to the minimal number of transpositions necessary to build it. For example, a single cycle permutation of length $k$ has degree $k-1$. The ring structure of the cohomology is then described as follows. Associate a function on the conjugacy class to each cohomology class. Convolute the functions, under the condition that a term  contributes if and only if the degrees of the permutations add. That convolution product on the class functions gives the structure constants in the cohomology ring of the Hilbert scheme of points on the plane \cite{LehnSorgerSymmetric}.

\subsubsection*{A First Example}
We will compute many structure constants in the following, and we start with a simple  example.  We denote the functions on the permutations in terms of the value the function $f$ takes at a given permutation times that permutation: $f=\sum_{\pi \in S_n} f(\pi) \pi$. 
 A first basic observation about the convolution product (\ref{Convolution}) is that if $f = \pi_1$ and $g = \pi_2$, then the convolution product at $\pi_1 \pi_2$ equals one, and is zero otherwise.
 Consider 
 a class function $\chi_{[\lambda]}$ on a conjugacy class $[\lambda]$:
 \begin{equation}
 \chi_{[\lambda]} = \sum_{\pi \in [\lambda]}  \pi \, .
 \label{ClassFunction}
 \end{equation}
 We  wish to partially compute the convolution product of class functions $\chi_{[n_1]} \ast \chi_{[n_2]}$. In particular, we calculate a coefficient of  interest, namely $(\chi_{[n_1]} \ast \chi_{[n_]})(\chi_{[n_1+n_2-1]})$ where we have dropped the cycles of length $1$ in the partition for the time being. We have chosen an example that satisfies the requirement that the degrees of the permutations add.
 
 Firstly, note that we have $n!/( (n-n_1)! n_1!) \times n_1! /n_1 
 =n!/((n-n_1)! n_1)$
 $n_1$-cycles. This is because we choose $n_1$ out of $n$ elements, and then decide on their order. However, cyclically reordering them gives the same $n_1$-cycle, hence the final division by the factor $n_1$. For a given $n_1$-cycle, to obtain a $(n_1+n_2-1)$-cycle, it is clear that we must have exactly one element in common between the $n_1$-cycle and the $n_2$-cycle. Thus, out of $n_2$ elements in the $n_2$-cycle, we pick one to be one out of $n_1$, and $n_2-1$ random out of the remaining $n-n_1$. We can choose to put the element that we chose out of the first $n_1$ first in the $n_2$-cycle, and we then have $(n_2-1)!$ inequivalent orderings for the other elements. Thus, we had $n_1 \times (n-n_1)!/( (n-n_1-n_2+1)! (n_2-1)!) \times (n_2-1)! = n_1 \times (n-n_1)!/(n-n_1-n_2+1)!$ options. Multiplying our options, we obtain a total of $n!/(n-n_1-n_2+1)!$ $(n_1+n_2-1)$-cycles.  Since there are a total of $n!/((n-n_1-n_2+1)!(n_1+n_2-1))$ $(n_1+n_2-1)$-cycles, each of those cycles obtains a coefficient $n_1+n_2-1$. We can write the final result in the  self-evident notation:
\begin{equation}
[1^{n-n_1} n_1] \ast [1^{n-n_2} n_2] = (n_1+n_2-1) [1^{n-n_1-n_2+1}n_1+n_2-1] + \dots \label{FirstStructureConstant}
\end{equation}
 Thus, the structure constant for the multiplication of the corresponding cohomology elements is $n_1+n_2-1$, since the degrees of the permutations add. Calculating the structure constant is an exercise in combinatorics. We have performed a simple such exercise and will encounter more intricate examples in due course.

\subsection{The Topological Conformal Field Theory}
In the previous subsection, we have reviewed the mathematical description of the cohomology ring of the Hilbert scheme of points on the complex plane. In this subsection, we want to describe the relation between the mathematics and the topological conformal field theory on the symmetric orbifold of the complex plane. The Hilbert-Chow projection of the Hilbert scheme of points maps the Hilbert scheme onto the symmetric orbifold space. The latter can be thought off as the target space of the symmetric orbifold conformal field theory. However, the conformal field theory, similarly to the scheme, desingularizes the target space \cite{Chen:2000cy,Ruan:2000mz}. Twisted sectors are necessarily added to the theory and they capture new directions in the configuration space  that render the theory consistent \cite{Dixon:1985jw,Dixon:1986jc}.

\subsubsection{The Conformal Field Theory}
Let us describe the field content of the seed conformal field theory on $\mathbb{C}^2$. We consider a $N=(4,4)$ supersymmetric conformal field theory in two dimensions with four real scalars $X^i$ and four Majorana fermions $\psi^i$. The complex scalars $X=(X^1+i X^2)/\sqrt{2}$ and $Y=(X^3+i X^4)/\sqrt{2}$ parameterize the complex two-plane $\mathbb{C}^2$. The fermions $\psi_X^\pm = (\psi^1 \pm i \psi^2)/\sqrt{2}$ and $\psi_Y^\pm = (\psi^3 \pm i \psi^4)/\sqrt{2}$ live in the tangent bundle. The (left and right) fermions can be bosonized: $\psi_X^\pm = e^{\pm i H_X}$ and $\psi_Y^\pm = e^{\pm i H_Y}$. The theory enjoys a $N=4$ superconformal symmetry in both the left- and right-moving sector with central charge $c=6$. The symmetry of the seed and orbifold theory is directly related to the hyperk\"ahler geometry of the target space.

The superconformal field theory has a continuous spectrum and the target space is non-compact. By picking a complex structure (or a preferred $N=2$ superconformal subalgebra), we can define a chiral ring of operators. The superconformal generators of the $N=2$ superconformal algebra can be chosen to be -- see e.g.  \cite{Polchinski2}:
\begin{eqnarray}
G^+ &=& i \sqrt{2} ( \psi_X^+ \partial {X} + \psi_Y^+ \partial {Y})
\nonumber \\
G^- &=& i \sqrt{2} ( \psi_X^- \partial \bar{X}  + \psi_Y^- \partial \bar{Y}) \, .
\end{eqnarray}
When we compute the chiral ring cohomology, we need to decide in which space we compute it. Let us consider the space of polynomials in the scalar  as well as the fermion fields. We moreover concentrate on chiral primaries such that we need to consider polynomial combinations of these fields only, and not their derivatives. Note that the operators $\bar{X}$,$\bar{Y}$ and $\psi^-_{X,Y}$ are not annihilated by the cohomological operator $\oint G^+$, while the operators $\psi^+_{X,Y}$ are exact. Thus, we are left with an operator ring of chiral primaries generated by the complex fields $X,Y$. These operators have non-singular operator products -- the  logarithmic singularity that condemns these operators to play a marginal role in the description of the conformal field theory cancels in all calculations inside the ring. The chiral operator ring that we defined coincides with the polynomial ring $\mathbb{C}[X,Y]$ which was the starting point for the description of the Hilbert scheme.

The de Rham cohomology of the complex plane, on the other hand, is concentrated in degree zero, and it is of dimension one. The cohomology coincides with the cohomology of the $N=1$ superconformal generator:
\begin{equation}
\oint  G = \oint  \delta_{ij} \psi^i \partial X^j \, ,
\end{equation}
along with its right-moving counterpart.
The charge $\oint G$ acts as the differential operator $d$ on differential forms represented as polynomials in the $\psi^i$ differentials, depending on coefficients which are functions of the coordinates $X^i$. Thus, the $\oint G$ cohomology  can be represented by the unit operator $1$ only.\footnote{We note that for the non-compact manifold at hand, the de Rham and the Dolbeault cohomology do not  coincide.}

The symmetric product conformal field theory  $Sym_N(\mathbb{C}^2)$ is defined by taking the tensor product of $n$ copies of the $N=(4,4)$ conformal field theory and dividing by the symmetric group $S_n$ that permutes the copies.   There is a list of definitions and prescriptions that determines the spectrum and correlation functions of the conformal field theory uniquely  \cite{Dixon:1985jw,Dixon:1986jc}. We wish to explain why the topologically twisted orbifold conformal field theory gives rise to the same cohomology ring as the  de Rham cohomology ring of the Hilbert scheme of points equipped with the cup product.

\subsubsection{The Cohomology}
Firstly, we want to match the operators that span the cohomology rings. The relevant operators have been described in the literature in great detail \cite{Lunin:2001pw}. The single cycle elements of the cohomology ring of the Hilbert scheme map onto certain  elements of the chiral ring in the orbifold twisted sectors.
They are denoted as the operators  $\sigma^{--}_{(n_1)}$ in \cite{Lunin:2001pw} and
can be described in terms of the twist operators $\tau_{(n_1)}$ of the lowest conformal dimension 
\begin{equation}
h (\tau_{(n_1)}) = \frac{c}{24} (n_1-\frac{1}{n_1})
\end{equation}
in the single cycle twisted sector, combined with an exponential in the bosonized fermions:
\begin{equation}
\sigma_{(n_1)}^{--} = e^{i \sum_{I=1}^{n_1} \frac{n_1-1}{2n_1} (H_X^I + H_Y^I)}
\tau_{(n_1)} \, .
\end{equation}
The fermions give total $U(1)_R$ charge $q=n_1-1$ to the operator which is of conformal dimension
\begin{equation}
h=\frac{1}{4}(n_1-\frac{1}{n_1})+ \frac{n_1}{4} \frac{(n_1-1)^2}{n_1^2} = \frac{n_1-1}{2}=\frac{q}{2}  \, ,
\end{equation}
and similarly in the right-moving sector. We have assumed that $n_1$ out of $n$ copies of the symmetric product are involved in the operator.  Permutations that consist of multiple cycles correspond to non-singular multiplications of operators $\sigma_{(n_i)}^{--}$. We still need to render the operator gauge invariant by conjugating with permutations. For each given permutation conjugacy class, we have precisely one chiral-chiral ring element of the type we discussed. These represent the cohomology elements one to one. While other chiral primaries exist, they will be trivial in de Rham cohomology, similarly to what we saw above for the seed theory. Indeed, if the original (de Rham) cohomology contains a single identity operator, the second quantized string perspective of \cite{Dijkgraaf:1996xw} or the mathematics result \cite{Goettsche} shows that the de Rham cohomology of the symmetric product conformal field theory is captured by the Poincar\'e polynomial (\ref{PoincarePolynomialPlane}). 
The crucial question becomes whether the operator product of these operators in the chiral ring agrees with the cup product of cohomology elements described in subsection \ref{CupProductComplexPlane}. 
\subsubsection{The Chiral Ring}
In the chiral ring, the product of operators can only be non-zero when R-charge conservation is satisfied. Moreover, two operators that are in twisted sectors labelled by the permutations $\pi_1$ and $\pi_2$ respectively, produce an operator in the twisted sector labelled by $\pi_1 \pi_2$. Consider for starters two operators in twisted sectors corresponding to permutations that consist of a single cycle $\pi_1=(n_1)$ and $\pi_2=(n_2)$. Their R-charges are  $n_1-1$ and $n_2-1$. Thus, if they produce a third operator with a single cycle, it must have length $n_1+n_2-1$. The operator product coefficient of the operators ${\sigma}_{(n_i)}^{--}$ was computed in \cite{Lunin:2001pw} and equals 
\begin{equation}
|C^{---}_{n_1,n_2,n_1+n_2-1}|^2 = \frac{n_1+n_2-1}{n_1 n_2}
\end{equation}
If we normalize 
\begin{equation}
\sigma_{(n_i)} = n_i {\sigma}_{(n_i)}^{--} \, ,
\end{equation} 
the operator product produces precisely the corresponding operator $\sigma_{(n_1+n_2-1)}$ with pre-factor equal to one.
This fact should be compared to the basic observation above equation (\ref{ClassFunction}). Finally, we note that to preserve R-charge we can allow for at most zero or one elements to overlap when taking the product of individual cycles. When we have zero overlap, the product is trivial. When we have an overlap of one, the product is as above.

Next, consider the generic case of a product of operators $\sigma_1 = \sigma_{(n_1^1)} \sigma_{ (n_2^1)} \sigma_{{(n_3^1)}} \dots$ with $\sigma_2= \sigma_{(n_1^2)} \sigma_{ (n_2^2)} \dots$. The  $U(1)_R$ charge of the initial operators is $\sum_{i} (n_i^a-1)$. This is the total number of transpositions in the initial twisted sector permutation labels. Charge conservation  guarantees that the total number of transpositions is conserved in the operator product expansion. Moreover, permutations that label twisted sectors compose (uniquely), matching the basic observation above equation  (\ref{ClassFunction}). Finally, gauge invariant operators are conjugation invariant and therefore correspond to class functions on the permutation group. Thus, we reproduce  the convolution product (\ref{Convolution}) with the additional requirement of degree conservation from the product of the chiral ring operators in the symmetric orbifold conformal field theory.
This is in full accord with an alternative route, which is to mathematically abstract \cite{Chen:2000cy,Ruan:2000mz} the cohomology coded in orbifold conformal field theory \cite{Dixon:1985jw,Dixon:1986jc}, to then prove that it is   equivalent to the Hilbert scheme cohomology \cite{FantechiGoettscheOrbifoldCohomology}.

\subsection{A Plethora of Results on the Cohomology Ring}
\label{Plethora}
We have established that calculations in the cohomology ring of the symmetric orbifold conformal field theory on the complex two-plane $\mathbb{C}^2$ reduce to calculations in the symmetric group $S_n$. The latter can be notoriously hard, and it is a domain of mathematics in itself to push the frontiers of what we can concretely compute in permutation groups and which combinatorics we can compactly count. We believe it is useful to review a  bit of  combinatorial knowledge relevant to the cohomology ring structure constants. We start by recalling what is known about the connection coefficients (i.e. structure constants) of conjugacy classes in general, and then restrict to the case where they satisfy the additive degree condition or equivalently, R-charge conservation.\footnote{One reason for discussing the combinatorics more generally  is that the other connection coefficients are relevant to models beyond the complex two-plane $\mathbb{C}^2$ on which we concentrate in this section.}
\subsubsection{The Product of Conjugacy Classes of $S_3$ and $S_4$}
The permutation group of one element is trivial. The permutation group of two elements $S_2=\mathbb{Z}_2$ has two elements, each in its conjugacy class, and these trivially compose. Only the composition $(12)(12)=1$ violates the degree condition.

The first slightly non-trivial case is the permutation group $S_3$. It contains $3!=6$ elements, and three conjugacy classes denoted $[1^3],[12],[3]$ in terms of partitions of three. The first has one element, the second three, and the third two. The multiplication table for $S_3$ is 
\begin{center}
\begin{tabular}{|c|c|c|c|c|c|c|}
\hline
\text{Element} & () &	(1 2)&	(2 3)&	(1 3)&	(1 2 3)&	(1 3 2) \\
\hline
()&	()	&(1 2)	&(2 3)&	(1 3)&	(1 2 3)&	(1 3 2) \\
\hline
(1 2)&	(1 2)&	()&	(1 2 3)&	(1 3 2)&	(2 3)&	(1 3) \\
\hline
(2 3)&	(2 3)&	(1 3 2)&	()&	(1 2 3)&	(1 3)&	(1 2)\\
\hline
(1 3)&	(1 3)&	(1 2 3)&	(1 3 2)&	()&	(1 2)&	(2 3)\\
\hline
(1 2 3)&	(1 2 3)&	(1 3)&	(1 2)&	(2 3)&	(1 3 2)&	()\\
\hline
(1 3 2)&	(1 3 2)&	(2 3)&	(1 3)&	(1 2)&	()&	(1 2 3) \\
\hline
\end{tabular}
\end{center}
where we act first with the column element, and then with the row element.
The convolution algebra on conjugacy classes is 
\begin{center}
\begin{tabular}{|c|c|c|c|}
\hline
\text{Conj. Class} &	[()] &	[(12)] &	[(123)] \\
\hline 
[()] &	[()]&	[(12)]&	[(123)] \\
\hline 
[(12)] &	[(12)]&	3[()] + 3[(123)]&	2[(12)] \\
\hline
[(123)]	&[(123)]&	2[(12)]&	2[()] + [(123)] \\
\hline
\end{tabular}
\end{center}
As an example, consider the conjugacy class of transpositions $[(12)]$ and compose it with the conjugacy class of transpositions $[(12)]$. The conjugacy classes contain three elements each, and we convolute their sums. We obtain nine terms, three of which are the identity, and six of which are cyclic permutations of order three (of which there are two). This result is indicated in the third row, third column of the table. 

In the symmetric group $S_4$, one has permutations that are the product of two independent non-trivial cycles.\footnote{We used that $2+2=4$.}
The multiplication table for $S_4$  already becomes a bit cumbersome. We can still easily list the five conjugacy classes in partition or cycle notation: $[1^4]=[()],[1^2 2]=[(12)],[2^2]=[(12)(34)],[13]=[(123)],[4]=[(1234)]$. These have $1,6,3,8$ and $6$ elements respectively. Their convolution algebra is:
\begin{center}
\begin{tabular}{|c|c|c|c|c|c|}
\hline
\text{Conj. Class} &	[()] &	[(12)] &	[(12)(34)] & [(123)] &[(1234)] \\
\hline 
[()] &	[()] &	[(12)] &	[(12)(34)] & [(123)] &[(1234)]  \\
\hline 
[(12)] &	 	[(12)] &   6 [()]+ 2 [(12)(34)]+3 [(123)] & \dots & \dots &  \dots   \\
\hline
[(12)(34)]	&  [(12)(34)] & [(12)]+2 [(1234)] &\dots &\dots &\dots   \\
\hline
[(123)]&  [(123)]  & 4[(12)]+4 [(1234)] &\dots &\dots & \dots   \\
\hline 
[(1234)] &  [(1234)] & 4 [(12)(34)]+3 [(123)] & \dots& \dots&  \dots \\
\hline
\end{tabular}
\end{center}
where the remaining entries in the table are still fairly straightforward to compute. Indeed, these data have been compiled in Appendix I.B of \cite{JamesKerber}  for $S_{n \le 8}$. They can also be reproduced using the symbolic manipulation program GAP \cite{GAP}. Each coefficient in these tables that satisfies the degree condition is a finite $n$ structure constant of the cohomology ring.

\subsubsection{The Independence of the Order of the Group}
\label{nIndependence1}
Importantly,  when we take a particular perspective, the structure constants are independent of the order of the permutation group, as long as the order is large enough  \cite{IvanovKerov}.
Let us illustrate this theorem on a simple example. We have class convolution formulas for $n=2,3,4,5,6$:
\begin{eqnarray}
[2] \ast [2] &=& 1 [1^2]
\nonumber \\
{ [}12] \ast [12]&=& 3 [1^3] + 3 [3]
\nonumber \\
{[}1^22] \ast [1^22] &=& 6 [1^4] + 2 [2^2]+ 3 [13]
\nonumber \\
{[}1^32] \ast [1^32] &=& 10 [1^5] + 2 [1 2^2] + 3 [1^23]
\nonumber \\
{[}1^42] \ast [1^42] &=& 15 [1^6] + 2 [1^2 2^2] + 3 [1^33] \, . \label{NormalizationExample}
\end{eqnarray}
Consider the renormalized class sums:
\begin{equation}
A[\lambda] = \left( \begin{array}{c} n-r+m_1(\lambda) \\ m_1(\lambda) \end{array} \right) [\lambda] \, , \label{RenormalizedClassSum}
\end{equation}
where we first pick a value  $n$ for the group $S_n$ in which we embed our partitions of the positive integer $r$ by adding $1$'s as necessary. The original partition of the integer $r$ has $m_1(\lambda)$ $1$'s to start with. Consider the left hand side of (\ref{NormalizationExample}). Our original partition is $[2]$ and $r=2$. Since $m_1(\lambda)=0$, we do not renormalize the left hand side characters at all. On the right hand side, we need to renormalize. For instance, when we embed the $[1^2]$ partition into the $[1^3]$ partition, we need to pick $2$ out of $3-2+2=3$ elements, and we pick up a factor of $3$.  For the next lines, we need to pick two out of 4,5,6 et cetera. For the other terms, the renormalization factors are  trivial. We see that when we write:
\begin{eqnarray}
A[2] \ast A[2] &=&  A[1^2]  
\nonumber \\
A[12] \ast A[12]&=& A [1^3] + 3 A [3]
\nonumber \\
A[1^22] \ast A[1^22] &=& A [1^4] + 2 A [2^2]+ 3 A [13]
\nonumber \\
A[1^32] \ast A[1^32] &=& A [1^5] + 2 A [1 2^2] + 3 A [1^23]
\nonumber \\
A[1^42] \ast A[1^42] &=& A [1^6] + 2  A[1^2 2^2] + 3  A [1^33] \, , \label{NormalizationExample2}
\end{eqnarray}
the structure constants stabilize at the order which is the sum of the orders of the original partitions (namely four). This is generically true \cite{IvanovKerov}, and it shows that the convolution of symmetric group class functions is independent of the order of the group (at finite order, as long as it is large). Thus, taking the large $n$ limit on the formulas for these renormalized structure constants is trivial -- they do not depend on $n$ at large $n$.

\subsubsection{Partially General Results}
While there is no general closed form expression that is an efficient rewriting of the original combinatorial problem of determining the multiplication of conjugacy classes, there are  partially general and compact results for the connection coefficients.\footnote{See e.g. the introduction of \cite{Goulden}  for an overview.} Especially regarding single cycle permutations and transpositions, there are more compact formulas. Let us for instance mention that the multiplication of the $n$-cycle conjugacy class with any conjugacy class has a reasonably compact expression \cite{Goupil}. Here we concentrate on the top connection coefficients, namely those connection coefficients that satisfy the R-charge conservation condition:
\begin{equation}
n - l (\pi_1) + n - l(\pi_2) = n - l(\pi_3) \label{TopCondition}
\end{equation}
where $l(\pi_i)$ is still the length of the permutation $\pi_i$ which equals the number of cycles that make up the permutation (or the number of parts of the corresponding partition). 
When the structure constant labels satisfy this condition, the connection coefficients in the symmetric group are called top coefficients (since the condition (\ref{TopCondition}) is extremal). Determining the top connection coefficients of the symmetric group is a hard and interesting problem about which many partial results are known. 

Let us review such a result. Suppose we have two conjugacy classes labelled by $[\pi_{1,2}]$, and we wish to know the structure constant multiplying $[\pi_3]$, then we denote the number $c^{[\pi_3]}_{[\pi_1] [\pi_2]}$. This number is symmetric in the lower indices and satisfies the property
$h_{\pi_3} c^{[\pi_3]}_{[\pi_1] [\pi_2]}=
h_{\pi_2}  c^{[\pi_2]}_{[\pi_1] [\pi_3]}$ where $h_{\pi_i}$ is the number of elements in the conjugacy class $[\pi_i]$.
In \cite{FarahatHigman,GoupilBedard,GouldenJacksonCombinatorial}, the top connection coefficients for two permutations that multiply into a single cycle were determined.
Let the permutation conjugacy classes $[\pi_1]= [1^{i_1} 2^{i_2} \dots]$ and $[\pi_2] = [1^{k_1} \dots]$ correspond to two partitions of $n=n_3$  of lengths $l_1 = l(\pi_1) = i_1 + i_2 + \dots$ and $l_2=l(\pi_2)= k_1 + k_2 + \dots$. With the degree restriction $l_1 + l_2 = n_3 + 1$ we have the top connection coefficient 
\begin{equation}
c_{[\pi_1],[\pi_2]}^{[(n_3)]} = n_3 \frac{(l_1-1)! (l_2-1)!}{i_1! i_2! \dots
k_1! k_2! \dots} \, . \label{AGeneralStructureConstant}
\end{equation}
This result has a direct connection to two-coloured plane rooted-trees on $n_3$ edges \cite{GouldenJacksonCombinatorial} and has  been further understood using Lagrange's inversion theorem and Macdonald's theory of symmetric functions in \cite{GouldenJacksonSymmetric}. It gives an idea of the power of combinatorial theorems in the context of computing structure constants of the cohomology ring.

For illustration purposes, we apply the result (\ref{AGeneralStructureConstant}) to our favorite case once more, namely $[\pi_i]=[(n_i)]$ with $n_3=n$. We find the partition $i_1=n-n_1$ and $i_{n_1}=1$ with length $l_1=n-n_1+1$ and similarly the partition
$k_1=n-n_2$, $k_{n_2}=1$ with length $l_2=n-n_2+1$.  We obtain once more the structure constant (\ref{FirstStructureConstant}):
\begin{equation}
c_{[(n_1)],[(n_2)]}^{[(n_3)]} = n_1+n_2-1 \, ,
\end{equation}
but now as a very simple example of a much more general formula.
We stress that combinatorics theorems like (\ref{AGeneralStructureConstant}) provide new results on operator product expansions of chiral operators in symmetric orbifold conformal field theories. The underlying maps to e.g. rooted trees or cacti \cite{GouldenJacksonCombinatorial} provide an efficient diagrammatics for the operator product expansions.

\subsection{The Interaction}
It is interesting to delve a little deeper into how the isomorphism between the Hilbert scheme cohomology ring and the degree preserving convolution ring of class functions on the symmetric group is established \cite{LehnSorgerSymmetric,Vasserot}. In both rings one can identify a basic building block for the product which is the multiplication by the conjugacy class of transpositions (or its appropriate algebraic geometric dual).
In fact, there is an interesting graded poset structure on the space of conjugacy classes of $S_n$ which can be used to compute structure constants of the ring \cite{GoupilBedard}. 
 We reach the next level in the poset by composing with transpositions. (See Figure 2 of \cite{GoupilBedard} for a neat illustration.)
This structure  played a  role in the  construction
of a differential operator that cuts and joins cycles through multiplication by transpositions \cite{Goulden}. Firstly, one associates polynomials to conjugacy classes of permutations
\begin{equation}
\Phi(\pi) = p_{[\pi]}
\end{equation}
where $p_\lambda = p_{\lambda_1} \dots p_{\lambda_i}$ is a product of  power sum symmetric functions\footnote{We have $p_0=1$ and $p_{\lambda_i}=x_1^{\lambda_i}+\dots + x_n^{\lambda_i}$.} associated to a partition $\lambda$  equal to the cycle distribution $[\pi]$ of the permutation $\pi$. We then have a differential operator $H_{int}$ which satisfies \cite{Goulden}:
\begin{equation}
\Phi([1^{n-2}2] \ast \pi) = H_{int} \Phi(\pi) \, 
\end{equation}
for all permutations $\pi$.
The differential operator $H_{int}$ equals \cite{Goulden}:
\begin{equation}
H_{int} = \frac{1}{2} \sum_{i,j\ge 1} ij p_{i+j} \partial_{p_i} \partial_{p_j} + (i+j) p_i p_j \partial_{p_{i+j}} \, . \label{CutAndJoin}
\end{equation}
The first term in the operator joins two cycles while the second term cuts one cycle into two. This is the most elementary (cubic) interaction term in the cutting and joining of the second quantized string theory described in \cite{Dijkgraaf:1996xw,Jevicki:1998bm,Dijkgraaf:2003nw}, and it is familiar from (topological) string theory.
In \cite{LehnSorgerSymmetric}, the operator $H_{int}$ is restricted to interactions that preserve the degree. It is not hard to see that the only terms that respect the degree are the terms of the first (join) type. Using the transposition as a stepping stone, it is shown in \cite{LehnSorgerSymmetric} that the alternating character generates the whole ring (both on the combinatorial and the algebraic geometric side) and establishes an isomorphism between the top degree class multiplication and the Hilbert scheme cohomology ring. The (cutting and) joining method of proof is akin to reasonings in second quantized string theory, matrix string theory or two-dimensional topological gravity.

\subsection{The Structure Constants are Hurwitz Numbers}

In this subsection, we wish to recall the relation between the Hilbert scheme of the plane, the counting of permutations and the Hurwitz numbers \cite{LQW2005}.
We define the Hurwitz numbers  $H_n^{\mathbb{P}^1}(\lambda_0,\dots,\lambda_k)$ as the number of $(k+1)$ tuples of permutations $(\pi_0,\pi_1,\dots,\pi_k)$ in $S_n$ such that $\pi_i$ is of cycle type $\lambda_i$ and $\pi_0 \pi_1 \dots \pi_k=1$,  divided by $n!$. This counts the number of (possibly disconnected) covers of the sphere with $k+1$ branching points of branching types $\lambda_i$ \cite{Hurwitz}. 
When we restrict to counting only those covers that satisfy the additive degree relation, then we impose that 
\begin{equation}
n - l (\lambda_0)  = \sum_{i=1}^k (n- l(\lambda_i)) \, ,
\end{equation}
where $l(\lambda_i)$ are the lengths of the partitions $\lambda_i$. 
For ease of notation, we introduce the normalization factor $\zeta(\lambda)=\zeta([1^{m_1} 2^{m_2} \dots])$:
\begin{equation}
\zeta({\lambda}) = \prod_{i \ge 1} (i^{m_i} m_i!) \, .
\end{equation}
We then have, following \cite{LQW2005}, that the generalized structure constant $c_{\lambda_1 \dots \lambda_k}^{\lambda_0}$  is  the (top) coefficient of the conjugacy class $[\lambda_0]$ in the product of conjugacy classes $[\lambda_i]$:
\begin{equation}
 \prod_{i=1}^k [\lambda_i] = c^{\lambda_0}_{\lambda_1 \dots \lambda_k} [\lambda_0] + \dots
\end{equation}
We rewrite this by using the number of elements in the  given conjugacy class $[\lambda_0]$ and find:
\begin{eqnarray}
c^{\lambda_0}_{\lambda_1 \dots \lambda_k}  &=& \frac{\zeta_{\lambda_0}}{n!} c_{\lambda_0 \dots \lambda_k}^{[1^n]} 
\nonumber \\
&=& \zeta_{\lambda_0} H_n^{\mathbb{P}^1}(\lambda^0,\dots, \lambda^k) \, ,
\label{HurwitzNumbers}
\end{eqnarray}
where the final equation follows by the property of the Hurwitz numbers reviewed above. Thus, top connection coefficients for the multiplication of any number of conjugacy classes are equal to the Hurwitz numbers that count the number of (possible disconnected) branched covers of the sphere with the appropriate extremal branching behaviour. In the end, this is a direct consequence of description of the cohomology ring in terms of the symmetric group class functions \cite{LehnSorgerSymmetric}

\subsection{The Broader Context and a Proof of a Conjecture}

Let us put the results we obtained in a broader context.
The first remark we wish to make is that we studied a ring which is a canonical quotient ring of the cohomology ring for a Hilbert scheme of points on a (quasi-projective) surface $M$.
Indeed, the latter cohomology ring has an ideal generated by all operators that have a factor that takes a non-trivial value (i.e. not the identity) in the seed cohomology ring of $M$. If we divide the symmetric orbifold cohomology ring by that ideal, the quotient ring is isomorphic to the cohomology ring of the Hilbert scheme of the complex plane \cite{LQW2004}. Thus, the latter captures part of the ring structure for any manifold $M$. 

The second remark is that up till now, we have restricted ourselves to an analysis of the operator algebra and its structure constants. This is because in the non-compact case, the topological orbifold conformal field theory has no natural seed two-point function -- in the compact case the two-point function is given by a (finite) integral over the compact manifold $M$. 

Nevertheless, we can at this stage make a useful connection with the physical extremal correlators computed for three- and four-point functions in \cite{Pakman:2009ab} and conjectured for any number of operator insertions  in equation (4.58) of  \cite{Pakman:2009ab}.\footnote{See also the end of section 5 of \cite{Rastelli:2005ph} for a discussion.} In the physical theory, there are diagonalizable two-point functions. These can be used to trivially lower an index on the structure constants of the cohomology  ring (\ref{HurwitzNumbers}). This simple remark shows that the Hurwitz number structure constants (\ref{HurwitzNumbers}) coincide with the correlators (4.58) in \cite{Pakman:2009ab} when they pertain to the identity cohomology element, and proves the conjecture in that case. In fact, we make the sub-statement considerably more powerful in that we showed that the statement is valid at  large enough but finite order $n$, with appropriate normalizations of the operators (as in (\ref{RenormalizedClassSum})), and moreover extends to a large class of surfaces $M$.

\subsection{Summary and Lessons}

We briefly summarize this section and prepare the ground for the next one.
When we divide the cohomology ring of the Hilbert scheme of points on a surface by the ideal  generated by all non-zero elements in the cohomology of the original manifold, we obtain the cohomology ring of the Hilbert scheme of points of the plane. The cohomology ring of the plane has a cup product which agrees with the top coefficients of the convolution of the conjugacy classes in the symmetric group $S_n$. The convolution product counts permutations in given conjugacy classes that multiply to one. The Hurwitz numbers as well count these permutations. Therefore, the extremal operator product expansion, which satisfies the condition that they correspond to top coefficients, agree with the Hurwitz numbers. This statement is true at order $n$ larger than the sum of the size of the permutations involved in the operator product, and the structure constants are independent of $n$ at large enough, but finite $n$.

We have studied a non-compact symmetric orbifold conformal field theory. The ring structure of the original manifold $\mathbb{C}^2$ is trivial, and this simplifies the cohomology ring of the symmetric orbifold. 
In the next section, we will work with a   compact complex surface $M$ and thus add to the mix both  a non-trivial seed cohomology ring, as well as a topological two-point function.

\section{Compact Surfaces}
\label{M}
\label{CompactSurfaces}
In the previous section, we discussed the cohomology ring of the Hilbert scheme of points on the complex two-plane $M=\mathbb{C}^2$ and how  the ring relates to a ring of chiral operators in the corresponding topological symmetric orbifold conformal field theory. In this section, we wish to extend  that analysis to the case where the  manifold $M$ is a  (quasi-projective) compact complex surface (with trivial canonical bundle), for example $M=K3$ or $M=T^4$.\footnote{Surfaces with non-trivial canonical class can also be studied using similar techniques, but here we concentrate on the case most frequently encountered in string theory.} We wish to review that the chiral ring of the topological symmetric orbifold conformal field theory has a neat mathematical description in terms of  the cohomology ring of the seed manifold $M$, as well as the combinatorics of permutations \cite{LehnSorgerCupProduct}. From that description, we  derive old and new correlators in the (topological) symmetric orbifold conformal field theory, as well as their generic characteristics.

\subsection{The Orbifold Frobenius Algebra}
\label{Frobenius}
The starting point of the construction in \cite{LehnSorgerCupProduct} is the cohomology ring of the manifold $M$, and a linear form $T$ on the ring. The latter is the integral on the compact manifold $M$.\footnote{From now on, will have in mind compact K\"ahler manifolds $M$. Note that in this case the Dolbeault cohomology is a refinement of the de Rham cohomology. Thus, for the compact case at hand, we no longer have to worry about the distinction between these cohomologies, as we did in section \ref{ComplexPlane}. The chiral ring will match perfectly onto the de Rham cohomology, both for the seed theory and for the Hilbert scheme.}
 The combined structure forms a graded Frobenius algebra $A$. The results of \cite{LehnSorgerCupProduct} then construct from this algebra, and the combinatorics of the symmetric group, a symmetric product Frobenius algebra $A^{[n]}$ that coincides with the chiral ring of the topological symmetric orbifold conformal field theory \cite{Chen:2000cy,Ruan:2000mz,FantechiGoettscheOrbifoldCohomology}. The construction of the algebra $A^{[n]}$ is lengthy -- we will provide only the gist of the construction and refer to \cite{LehnSorgerCupProduct} for further details.

\subsubsection{The Tensor Frobenius Algebra and Permutations}

Following \cite{LehnSorgerCupProduct} we start with a  graded commutative, associative algebra $A$ over $\mathbb{C}$, with a unit and a linear form $T$ such that the bilinear form $T(ab)$ induced by $T$ is non-degenerate.\footnote{We encourage the reader to already imagine that the graded associative algebra $A$ is the algebra of cohomology elements of the manifold $M$ 
and to think of the bi-linear form $T$ as the topological conformal field theory two-point function, given by integrating the wedge product of two cohomology elements over the manifold $M$.}
Our graded Frobenius algebra  is a finite dimensional graded vector space with grades ranging from $-d$ to $d$, with a multiplication of degree $d$ and a unity element (necessarily of degree $-d$). These grades are the degrees of the cohomology elements on a manifold of complex dimension $d$, shifted by $-d$.\footnote{Our surface has  dimension $d=2$. We will nevertheless keep the dimension $d$ general for a while.}${}^{,}$\footnote{The reader familiar with conformal field theory is invited to simultaneously think of the grade as the sum of the Ramond-Ramond sector $U(1)_R$ charges, and the shift $-d=-c/3$ as arising from spectral flow.} The linear form $T$ is of degree $-d$. 

The tensor product $A^{\otimes n}$ is a Frobenius algebra of degree $nd$. The product of tensor product elements is the (graded) tensor product of the individual products in the factors:
\begin{equation}
(a_1 \otimes \dots \otimes a_n).(b_1 \otimes \dots \otimes b_n)
= \epsilon(a,b) (a_1 b_1) \otimes \dots \otimes (a_n b_n) \, ,
\end{equation}
where the sign $\epsilon(a,b)$ keeps track of the number of odd elements (i.e. fermions) we had to exchange in order to perform the multiplication.
The bi-linear form on the tensor product algebra $A^{\otimes n}$ is defined by 
\begin{equation}
T(a_1 \otimes \dots \otimes a_n ) = T (a_1) \dots T(a_n) \, .
\label{InducedBilinearForm}
\end{equation}
It is of degree $-nd$ and is again non-degenerate.
The symmetric group acts on the tensor product factors and the action takes into account the grading of the factors. For each bijection $f:\{ 1,2,\dots , n\} \rightarrow I$, there is a canonical isomorphism $A^{\otimes n} \equiv A^{\otimes I}$. We will often implicitly make use of this in the following.\footnote{At this stage, we have the topological tensor product conformal field theory Hilbert space with an action of the permutation group, the seed ring structure and a two-point function that extends to the tensor product.}

Next, we define maps between tensor products of the original Frobenius algebra with a different number of factors. First, we reduce the number of factors. Suppose we have a partition $n=n_1 + n_2 + \dots + n_k$. We then define a ring homomorphism through multiplication:
\begin{equation}
\phi_{n ,k}: A^{\otimes n} \rightarrow A^{\otimes k}:
a_1 \otimes \dots \otimes a_n \mapsto  (a_1 a_2 \dots  a_{n_1})
\otimes \dots \otimes   (a_{n_1+ \dots +n_{k−1}+1} \dots a_n) \, .
\end{equation} 
This generalizes to a map $\phi^\ast$ associated to any surfective map of indices $\phi: I \rightarrow J$. Moreover, we have the adjoint map $\phi_\ast: A^{\otimes J} \rightarrow A^{\otimes I}$ with respect to the bilinear forms (\ref{InducedBilinearForm}) induced on the tensor products.

 A particular example of these maps that we will encounter is the multiplication $\phi_{2,1}=\Delta^\ast: A \otimes A \rightarrow A$ as well as its adjoint, the co-multiplication $\Delta_\ast : A \rightarrow A \otimes A$.   The image of the unit under the combined map is called the Euler class $e(A)$ of the algebra $A$. Let us compute the Euler class in more familiar terms. Consider a (graded) basis $e_i$ of the algebra $A$. We introduce an expression for  the structure constants ${c_{ij}}^k$ of the algebra $A$ as well as for the metric $\eta_{ij}$ that arises from the bilinear form $T$:
 \begin{eqnarray}
e_i e_j &=& {c_{ij}}^k e_k = (-1)^{ij} {c_{ji}}^k e_k
\nonumber \\
T(e_i e_j) &=& \eta_{ij} = (-1)^{ij} \eta_{ji} =(-1)^i \eta_{ji} = (-1)^j \eta_{ji} \, .
 \end{eqnarray}
We drew attention to (anti-)symmetry properties of the structure constants and the metric (exploiting an obvious notation for the grading of the basis elements). We have a unique element in the cohomology that is the unit element and it has grade $-d$. By Poincar\'e duality, there is a unique volume element of grade $+d$ that evaluates under the linear form $T$ to $T(\text{vol})=1$. The co-multiplication map maps the unity element to ${c^{ji}}_0 e_i \otimes e_j$ since it is the adjoint of the multiplication map captured by the structure constants. We denoted the unity basis element $e_0=1$ by the index zero. If we then multiply  $e_i$ with $e_j$, we obtain $ 1 \xrightarrow{\Delta^\ast \circ \Delta_\ast} {c^{ji}}_0 {c_{ij}}^k e_k
 = {c^{ji}}_0 c_{ij0} \, \text{vol} = \chi(A) \, \text{vol}$ where $\chi(A) =\sum_i (-1)^i dim(A^i) = \eta_{ij} \eta^{ji}$ is the Euler-Poincar\'e characteristic of the algebra $A$. We thus  map unity to the Euler class $e(A)=\chi(A) \, \text{vol}$.\footnote{For the four-torus, the Euler number $\chi(T^4)$ is zero, and therefore $e(A_{T^4})=0$. For the $K3$ surface, the Euler number equals $\chi(K3)=24$.} 
 For future purposes, we introduce the notation
 \begin{equation}
 e^{g} = \otimes_{i \in I} e^{ g(i)} \in A^{\otimes I}
 \end{equation}
 for the tensor product of Euler classes associated to
 a function $g: I \rightarrow \mathbb{N}_0$. 
Next, we need to add ingredients that will be necessary to treat the combinatorial aspects  of the symmetric orbifold theory. We recall that every permutation $\pi$ has a degree $|\pi|$ given by the minimal number of transpositions necessary to construct it. 
For any subgroup $H \subset S_n$ and an $H$-stable subset $B \subset \{ 1, 2, \dots , n \}$, we write $H \setminus B$ for the space of orbits in the set $B$ under the action of the group $H$. We note that the degree $|\pi|$ of a permutation can be identified as $|\pi| = n - |\langle \pi \rangle \setminus [n]|$ where $\langle \pi \rangle$ is the subgroup of $S_n$ generated by the permutation $\pi$. Indeed, this is the total number of elements minus the number of cycles, and thus the sum of cycle lengths minus one.
This notation allows us to define the graph defect $g$ for two permutations $\pi$ and $\rho$ evaluated on a set $B$:
\begin{equation}
g(\pi,\rho)(B) = \frac{1}{2} (|B|+2 -|\langle \pi \rangle \backslash B|-|\langle \rho \rangle \backslash B| - |\langle \pi \rho \rangle \backslash B|) \, . \label{GraphDefect}
\end{equation}
The graph defect is a positive integer that has an interpretation as the genus of a Riemann surface \cite{LehnSorgerCupProduct}. 
\subsubsection{The Twisted Sectors, the Multiplication and the Gauge Invariants}
We introduce two further algebras $A \{ S_n \}$ and $A^{[n]}$ \cite{LehnSorgerCupProduct}. The first is the analogue of the unprojected space of operators in the topological orbifold theory, while the second is the subspace of $S_n$ gauge invariants. We need to define the first space, an action of the permutation group on the space, and most importantly, the equivariant product in the algebra.
We define the algebra $A \{ S_n \}$ for starters as the (graded) vector space of tensor products of operators multiplying permutations:
\begin{equation}
A \{ S_n \} = \bigoplus_{\pi \in S_n} A^{\otimes \langle \pi \rangle \backslash [n]} \, \pi \, .
\end{equation}
Each orbit of the permutation is assigned a single operator.
The grading of an element $a \, \pi$ is the sum of the grades in the tensor algebra, namely $|a \, \pi| = |a|$. The action of the group $S_n$ on the set $\{ 1, 2, \dots, n \}=[n]$ gives rise to a bijection between the orbits of a permutation $\pi$ and its conjugates:
\begin{equation}
\sigma : \langle \pi \rangle  \backslash [n] \rightarrow \langle \sigma \pi \sigma^{-1} \rangle \backslash [n] : x \mapsto \sigma x \, .
\end{equation}
Thus we have an isomorphism $\tilde{\sigma}$
\begin{equation}
\tilde{\sigma} : A \{ S_n \} \rightarrow A \{ S_n \} : a \pi \mapsto \sigma (a) \sigma \pi \sigma^{-1} \, ,
\end{equation}
which defines the action of the symmetric group $S_n$ on the vector space $A \{ S_n \}$. 
The vector space $A^{[n]}$ is the subspace of symmetric group invariants $A^{[n]}=(A \{ S_n \})^{S_n}$. 

Finally, we come to the crucial definition of the product on $A \{ S_n \}$ \cite{LehnSorgerCupProduct}. Firstly, note that any inclusion of subgroups $H \subset K$ leads to a surjection of orbit spaces $H \setminus [n] \rightarrow K \setminus [n]$ and by the previous constructions to a map
\begin{equation}
f^{H,K} : A^{\otimes H \setminus [n]} \rightarrow A^{\otimes K \setminus [n]} \label{Plungers}
\end{equation}
and its adjoint $f_{K,H}$. These maps allow us to bring operators that lie in different twisted sectors into a common space in which we are able to multiply them, then to bring them back into the product twisted sector.
Indeed, we define the operator multiplication map $m_{\pi ,\rho}$ \cite{LehnSorgerCupProduct}:
\begin{equation}
m_{\pi,\rho}: A^{\otimes \langle \pi \rangle \setminus [n]}
\otimes A^{\otimes \langle \rho \rangle \setminus [n]}
\rightarrow A^{\otimes \langle \pi \rho \rangle \setminus [n]}
\end{equation}
by
\begin{equation}
m_{\pi,\rho}(a \otimes b) = f_{\langle \pi,\rho \rangle,\pi \rho} (f^{\pi,\langle \pi,\rho \rangle} (a) . f^{\rho,\langle \pi,\rho \rangle} (b) . e^{g (\pi,\rho)}) \, . \label{MultiplicationFormula}
\end{equation}
Note that the middle multiplication takes place in the tensor product associated to the subgroup $\langle \pi,\rho \rangle$ generated by both permutations $\pi,\rho$ separately.
The multiplication map $m_{\pi,\rho}$ not only prescribes how to sensibly combine operators in different tensor product factors, it also tells us that we need to take into account a correction factor that depends on the graph defect as well as on the Euler class $e(A)$ of the seed algebra $A$. It moreover prescribes how to co-multiply from the smaller set of orbits of the subgroup $\langle \pi,\rho \rangle$ to the orbits of the subgroup $\langle \pi \rho \rangle$. 
Finally, the cup product is given by:
\begin{equation}
a \pi . b \rho = m_{\pi,\rho} (a \otimes b) \pi \rho \, ,
\label{Multiplication}
\end{equation}
namely, twists compose.
The product is associative, $S_n$ equivariant and homogeneous of degree $nd$, as proven in \cite{LehnSorgerCupProduct}. In other words, we have $|a \pi \cdot b \rho|=|a \pi| + |b \rho| + nd$.

The equivariant ring structure on $A \{ S_n \}$ induces a ring structure on the space of invariants $A^{[n]}$ and moreover, the latter ring is a subring of the centre of $A \{ S_n \}$ \cite{LehnSorgerCupProduct}. Thus, elements (anti-)commute in the orbifold ring $A^{[n]}$, as expected. The algebra $A^{[n]}$ is again a graded Frobenius algebra.

\subsubsection*{Remarks}
\begin{itemize}

\item
Let us describe  briefly how the generic construction in this section relates to the case of the cohomology ring of the Hilbert scheme of the complex plane $\mathbb{C}^2$ in section \ref{ComplexPlane}. In that case, the seed Frobenius algebra $A$ is trivial. Permutations multiply as in the multiplication formula (\ref{Multiplication}). What remains to show is that the multiplication (\ref{Multiplication}) of degree $nd$ preserves the degree of the permutations. We will show that it does. 
The degree of the unit operator in $A$ is equal to $-d$. We moreover recall the relation between the degree of a permutation and the number of orbits:
\begin{equation}
| \pi| =  n - | \langle \pi \rangle \setminus [n]| \, .
\end{equation}
If we consider the element $1 \, \pi$ as an element of $A^{\otimes \langle \pi \rangle \setminus [n]} \, \pi$ then its degree is equal to $|\langle \pi \rangle \setminus [n]|$ times $-d$. Thus, if we have cohomology elements $a=1=b$, the degree condition for the product $m_{\pi,\rho}$ reduces to:
\begin{equation}
|1\pi \cdot 1\rho| %
=
 -d| \langle \pi \rangle \setminus [n]|-d| \langle \rho \rangle \setminus [n]|+d n
= d ( |\pi| + |\rho|) -dn = -d | \langle \pi \rho \rangle \setminus [n] |
\end{equation}
and therefore $|\pi \rho|= n - | \langle \pi \rho \rangle \setminus [n] |=|\pi| + |\rho|$. This is indeed the degree or R-charge conservation condition familiar from section \ref{ComplexPlane}. More generally, the  degree condition on the product agrees with total R-charge conservation in the conformal field theory.
\item An isomorphism between the algebra of gauge invariants $A^{[n]}$ and a Fock space as for a second quantized string  \cite{Dijkgraaf:1996xw} can be established \cite{Goettsche,LehnSorgerCupProduct}.
\item In \cite{LehnSorgerCupProduct} the cup product on Hilbert schemes of surfaces was proven to coincide with the efficient description reviewed above. Moreover, it was proven to coincide with the mathematical reformulation of the symmetric product cohomology \cite{Chen:2000cy,Ruan:2000mz,FantechiGoettscheOrbifoldCohomology}. 
\item
Note that since the Euler class $e(A({T^4}))$ of $T^4$ is zero, because the Euler number is, the correction associated to the Euler class is absent for the $T^4$ topological symmetric orbifold conformal field theory. Moreover, the multiplication map will set to zero any multiplication that has non-zero genus or graph defect. 
\item 
For the case of $M=K3$, the Euler class correction is important. 
\item Whenever the graph defect (or genus) of two permutations $\pi,\rho$  we multiply is larger or equal to two in a given orbit of the group $\langle \pi,\rho \rangle$ generated by the two permutations, then the multiplication $m_{\pi,\rho}$ contains a factor of $e(M)^2$ and is therefore equal to zero. In this sense, the multiplication formula contains one-loop corrections only. Provided we can identify the covering surface with the bulk string world sheet (see e.g. \cite{Pakman:2009zz}), there are at most  one loop corrections in the bulk topological string theory.
\item Finally, we summarize in geometric terms, familiar from generic two-dimensional topological field theories, the mechanics behind the results we found. Recall that these theories are solvable in terms of their (pants) three-point functions, combined with cutting and gluing. From the three-point functions, we can compute the operator that inserts a handle in a Riemann surface. We have computed the handle operator $e={c_{i}}^{ij} e_j$ to be (multiplication by) the Euler class. Since the Euler class squares to zero, the insertion of two handles makes for a vanishing amplitude.

\end{itemize}
We have reviewed the topological symmetric orbifold conformal field theory cohomology ring for a (quasi-projective) compact manifold $M$ (with trivial canonical class) \cite{LehnSorgerCupProduct}, and added some remarks that serve to guide the physics reader through the construction. We will next employ the efficient description to recompute a few structure constants and explicitly match them to the physics literature on the symmetric orbifold conformal field theory. These calculations confirm the general picture drawn above, as they must. 

\subsection{Cup Products of Single Cycle Elements}
\label{CupProductsSingleCycles}
We compute the operator product  of two operators  associated to two permutations that are each single cycles with either zero elements overlapping, one, or two. One reason for performing  these calculations is that they can be compared to calculations in the physical conformal field theory \cite{Jevicki:1998bm,Lunin:2000yv,Lunin:2001pw}. A second reason is that these correlators have been compared to bulk three-point functions \cite{Dabholkar:2007ey,Gaberdiel:2007vu} and thus provide a concrete holographic bridge to string theory in three-dimensional anti-de Sitter space.
\subsubsection{Without Overlap}
In the case where we consider two operators $\alpha \sigma_{(n_1)} $ and $\beta \sigma_{(n_2)}$ in the algebra $A \{ S_n \}$ concentrated in $n_i$ factors of the symmetric product $A^{\otimes n}$ with zero overlap in the respective factors, the product is trivial, namely it is the ordinary tensor product of the individual operators.\footnote{In the algebra of gauge invariants $A^{[n]}$ this situation will hardly occur. After making gauge invariants, various cycles are bound to overlap. In that circumstance, we have zero overlap for certain terms only.}${}^{,}$\footnote{In the following we will often use the notation $\alpha,\beta,\dots$ for operators in $A$, associated to a single orbit of the permutation. For example, we have $\alpha \sigma_{(n_1)} = \dots \otimes 1 \otimes \alpha \otimes 1 \otimes \dots  \sigma_{(n_1)}$ where $\alpha \in A$ corresponds to the orbit of $\sigma_{(n_1)}$ of length $n_1$.}

\subsubsection{A Single Overlap}
We discuss a second case of a product of operators $\alpha \pi$ and $\beta \rho$ in the algebra $A \{ S_n \}$, associated to two single cycle permutations $\pi = \sigma_{(n_1)}$, $\rho=\sigma_{(n_2)}$ with a single element in common.   As a result, we have the multiplication of permutations $\pi \rho = \sigma_{(n_1+n_2-1)}$ and we act on the minimal space of indices $\{1,2,3,\dots,n_1+n_2-1 \} \equiv [n_1+n_2-1]$.\footnote{This notation for a set conflicts with the notation for conjugacy classes. The context should be sufficient to distinguish the concepts.} To compute the product of such operators in the algebra  $A \{ S_n \}$ according to the rules laid out in subsection \ref{Frobenius}, we establish a few preliminaries. The space of orbits of the various subgroups generated by the  permutations is -- with mild abuses of notation, and a particular choice of permutation action --:
\begin{eqnarray}
\langle \pi \rangle \setminus [n_1+n_2-1] &=& \{ [n_1],n_1+1,n_1+2,\dots, n_1+n_2-1 \}
\nonumber \\
\langle \rho \rangle \setminus [n_1+n_2-1] &=& \{ 1,2,\dots,n_1-1, \{ n_1,n_1+1,\dots,n_1+n_2-1 \} \}
\nonumber \\
\langle \pi \rho \rangle \setminus [n_1+n_2-1] &=& \{ [n_1+n_2-1] \}
\nonumber \\
\langle \pi , \rho \rangle \setminus [n_1+n_2-1] &=& \{ [n_1+n_2-1] \} \, .
\end{eqnarray}
We have the graph defect $g$ (\ref{GraphDefect}) for the stable subset $B=[n_1+n_2-1]$:
\begin{eqnarray}
g(\pi,\rho)([n_1+n_2-1]) &=& \frac{1}{2}
(n_1+n_2-1+2-n_1-n_2-1)=0 \, . \label{FirstGraphDefect}
\end{eqnarray}
We then apply maps $f^{\pi,\langle \pi,\rho \rangle}$ and
$f^{\rho, \langle \pi, \rho \rangle}$  to  elements $\alpha$ and $\beta$ which are elements of the cohomology associated to the orbits of $\sigma_{(n_1)}$ and $\sigma_{(n_2)}$ respectively, and the identity otherwise, such that they become elements of a single algebra $A^{ \langle \pi,\rho \rangle \setminus [n_1+n_2-1]}$ associated to the orbit of the subgroup $\langle \sigma_{(n_1)},\sigma_{(n_2)} \rangle$. In this single factor, we  multiply $\alpha$ and $\beta$ according to formula (\ref{MultiplicationFormula}), and we receive no extra factor from the Euler class of the algebra $A$ since the graph defect $g$ (\ref{FirstGraphDefect}) is zero. 
Moreover, for this case, we have that the orbits of $\langle \pi \rho \rangle$ and $\langle \pi,\rho \rangle$ are the same, and thus that the map $f_{\langle \pi,\rho \rangle,\pi \rho}$ is trivial. Thus the product $m_{\pi,\rho}(\alpha \otimes \beta)= \alpha \beta$ is merely the multiplication of cohomology elements, which we combine with the non-trivial composition of permutations to obtain the final result for the multiplication in $A \{ S_n \}$:
\begin{equation}
\alpha \sigma_{(n_1)} \cdot \beta \sigma_{(n_2)} = \alpha \beta \sigma_{(n_1+n_2-1)} \, .
\end{equation}
To obtain a product in $A^{[n]}$ in which we concentrate on the right hand side on a single cycle permutation generated by a single overlap, we merely need to count elements in conjugacy classes, as we did in subsection \ref{CountingConjugacyClasses}. We therefore  find the structure constants of the original ring, captured by the product $\alpha \beta$, times the combinatorial coefficient $n_1+n_2-1$. Thus, we have computed a subset of three-point functions. In this example, the ring structure factorized into the seed ring structure times combinatorics.

\subsubsection{An Overlap of Two}
In the next example we study two operators multiplying single cycles, overlapping in two factors of the tensor product space. The composition of permutations on which we focus is
$\sigma_{(n_1)} \sigma_{(n_2)} = \sigma_{n_1+n_2-3}$
where we have an overlap of two entries in the cyclic permutations that are in reverse order and therefore generate a single longer cycle. We concentrate on the example $\pi=\sigma_{(n_1)} =(12 \dots n_1)$ 
and $\rho=\sigma_{(n_2)}=(2,1,n_1+1,n_1+2,\dots,n_1+n_2-2)$ which composes to $\pi \rho=\sigma_{n_1+n_2-3}=(1,n_1+1,n_1+1,\dots,n_1+n_2-2,3, \dots,n_1)$. We apply again the iron logic of \cite{LehnSorgerCupProduct} and first compute the orbits of the permutations in the set $\{ 1,2, \dots,n_1+n_2-2 \}$:
\begin{eqnarray}
\langle \pi \rangle \setminus [n_1+n_2-2] &=& \{ [n_1],n_1+1,n_1+2,\dots, n_1+n_2-2 \}
\nonumber \\
\langle \rho \rangle \setminus [n_1+n_2-2] &=& \{ 3,4,\dots,n_1, \{1,2,n_1+1,n_1+2,\dots,n_1+n_2-2 \} \}
\nonumber \\
\langle \pi \rho \rangle \setminus [n_1+n_2-2]&=& \{2, \{ 1,3,4,\dots \dots, n_1+n_2-2 \} \}
\nonumber \\
\langle \pi , \rho \rangle \setminus [n_1+n_2-2] &=& [n_1+n_2-2] \, ,
\end{eqnarray}
as well as the graph defect $g$:
\begin{eqnarray}
g(\pi,\rho)([n_1+n_2-2]) &=& \frac{1}{2}
(n_1+n_2-2+2-(n_1-1)-(n_2-1)-2)=0 \, ,
\end{eqnarray}
which is once again zero.
As in the previous subsection,  after projection and multiplication, we have the  product $\alpha \beta$ of the elements $\alpha$ and $\beta$ associated to the individual factors. However, now  we have a non-trivial map $f_{\langle \pi,\rho \rangle , \pi \rho}$ which maps from a space with a single factor $A$ back to the tensor product $A^{\otimes 2}$ associated to the two orbits of the product $\pi \rho$. This adjoint map back is nothing but the co-multiplication $\Delta_\ast$. In slightly more detail than before, we have the maps:
\begin{equation}
\Delta (e_i \otimes e_j) = f^{\pi \rho, \langle \pi,\rho \rangle} (e_i \otimes e_j) = e_i e_j = {c_{ij}}^k e_k
\end{equation}
as well as the adjoint condition:
\begin{equation}
T(e_l , f^{\pi \rho, \langle \pi,\rho \rangle} (e_i \otimes e_j) ) =  {c_{ij}}^k \eta_{lk} = (-1)^l c_{ijl}
= T( f_{ \langle \pi,\rho \rangle,\pi \rho} e_l , e_i \otimes e_j) \, .
\end{equation}
The unique solution is 
\begin{equation}
\Delta_\ast (e_l)=f_{ \langle \pi,\rho \rangle,\pi \rho} e_l
= (-1)^l {c^{mk }}_l e_k \otimes e_m \, ,
\end{equation}
since
\begin{eqnarray}
T( (-1)^l {c^{mk}}_l e_k \otimes e_m, e_i \otimes e_j) 
&=& (-1)^l {c^{mk}}_l (-1)^{mi} \eta_{ki} \eta_{mj} 
\nonumber \\
& =& (-1)^l
 c_{ijl}  \, .
\end{eqnarray}
Thus, we find the final product:
\begin{equation}
\alpha \sigma_{(n_1)} \cdot \beta \sigma_{(n_2)} = \Delta_\ast (\alpha \beta) \sigma_{(n_1+n_2-3)} \, ,
\end{equation}
which we can render explicit  in terms of the structure constants ${c_{ij}}^k$ and the metric $\eta_{ij}$ of the seed theory:
\begin{equation}
\Delta_\ast( \alpha^i \beta^j {c_{ij}}^k e_k )
= \alpha^i \beta^j {c_{ij}}^k (-1)^k {c^{nm}}_k  e_m \otimes e_n \, .
\end{equation}
Next, we wish to add in the gauge invariance combinatorics that arises when exploiting this result to obtain a structure constant in the orbifold algebra of gauge invariants $A^{[n]}$.
We   count the number of such   double overlap multiplications that appear when multiplying conjugacy invariant operators. We again start by picking $n_1$ elements out of $n$ elements (say), and ordering them in $(n_1-1)!$ ways, to obtain $n!/(n-n_1)!/n_1$ different operators $\sigma_{(n_1)}$. We pick two consecutive elements out of the $n_1$ elements, which we can do in $n_1$ ways -- we must put them in opposite order in the second permutation --, and moreover, we pick $(n_2-2)$ elements out of $n-n_1$, and order them in $(n_2-2)!$ ways, which we can do in $(n-n_1)!/(n-n_1-n_2+2)!$ ways.  Thus, this gives a total of $n! /  (n-n_1-n_2+2)!$ of combinations of the type we are looking for. We have 
$n!/ ( (n-n_1-n_2+3)! (n_1+n_2-3))$ elements in the conjugacy class of $\sigma_{n_1+n_2-3}$, and we then still have $n-n_1-n_2+3$ choices for the extra lonely element we need, for a total of 
$n!/((n-n_1-n_2+2)! (n_1+n_2-3))$ choices.  Thus we have a prefactor of 
\begin{equation}
n_1+n_2-3 \,  \label{CombinatorialFactor3}
\end{equation}
that arises from the combinatorics.
Apart from the structure constants and adjoint map that arises from the original cohomology ring, this is the structure constant of the multiplication of the classes of $\sigma_{(n_1)}$ with $\sigma_{(n_2)}$ into $\sigma_{(n_1+n_2-3)}$. If we denote the gauge invariant operator sum (without extra normalization factor) by  $O_{(n_i)}(\alpha)$, we find the operator product coefficient: 
\begin{equation}
O_{(n_1)}(\alpha) O_{(n_2)}(\beta)  = (n_1+n_2-3) O_{(1)(n_1+n_2-3)}
(\Delta_\ast(\alpha \beta)) + \dots
\end{equation}
where on the right hand side, we need to split the co-product over the two tensor factors associated to the cycle of length one and the cycle of length $n_1+n_2-3$.

Thus, we computed a few structure constants in the orbifold cohomology ring of the Hilbert schemes of points on $M=T^4$ and $M=K3$ among others. We wish to compare these results to operator product expansions of chiral primary operators in the physical symmetric orbifold conformal field theory. Those were obtained using rather different techniques.

\subsubsection{To Match the Physical Correlators}
We wish to confirm that the mathematical construction of subsection \ref{Frobenius}  indeed captures the physical operator product expansions that survive topological twisting.
From the references \cite{Jevicki:1998bm,Lunin:2000yv} and in particular \cite{Lunin:2001pw}, we glean a few structure constants in the symmetric orbifold conformal field theory chiral ring.  The appropriate extraction from \cite{Lunin:2001pw} is performed in Appendix \ref{LuninMathurExtraction}. 
As in \cite{Lunin:2001pw}, we use the notation $\sigma^{\pm,\pm}_{(n_i)}$ to indicate a chiral ring primary associated to the $(-,-)=(0,0)$ Dolbeault cohomology, the $(+,-)=(2,0)$, $(-,+)=(0,2)$ and the $(+,+)=(2,2)$ Dolbeault cohomology of the manifolds $M=K3$ or $M=T^4$. Here, we focus on these manifolds and classes only. Moreover, following \cite{Lunin:2001pw} we factorize the structure constants into a structure constant associated to two-dimensional massless left-movers and a structure constant associated to right-movers in the symmetric orbifold conformal field theory. The  structure constants from 
\cite{Lunin:2001pw} can then be summarized by -- see Appendix \ref{LuninMathurExtraction} :
\begin{eqnarray}
{C}^{1_{n_1},1_{n_2};1_{n_3}}_{n_1 n_2;n_3} &=& \left(
\frac{(1_{n_1} n_1+ 1_{n_2} n_2 +1_{n_3} n_3 +1)^2}{4 n_1 n_2 n_3}  
\right)^{\frac{1}{2}}
\label{NonReducedStructureConstant} 
\end{eqnarray}
where the individual operators carry $U(1)_R$ charge $n_i+1_{n_i}$ and $1_{n_i}$ is $\pm 1$ depending on the (left or right) upper index of the operator $\sigma^{\pm,\pm}_{n_i}$, and we assume the R-charge constraint $n_1+1_{n_1} + n_2+1_{n_2} = n_3+ 1_{n_3}$ \cite{Lunin:2001pw}.
The (left or right) structure constants that preserve $U(1)_R$ charge are the cases \cite{Jevicki:1998bm}:
\begin{eqnarray}
C^{---}_{n_1,n_2,n_1+n_2-1} &=& ( \frac{n_1+n_2-1}{n_1 n_2})^{\frac{1}{2}}
\nonumber \\
C^{-++}_{n_1,n_2,n_1+n_2-1} &=& ( \frac{n_2}{n_1(n_1+n_2-1)})^{\frac{1}{2}} \nonumber \\
C^{--+}_{n_1,n_2,n_1+n_2-3} &=& (\frac{1}{n_1 n_2(n_1+n_2-3) })^{\frac{1}{2}} \, ,
\end{eqnarray} and these chiral structure constants lead to the non-chiral operator product terms:
\begin{eqnarray}
\sigma_{n_1}^{--} \sigma^{--}_{n_2} &=& \frac{n_1+n_2-1}{n_1 n_2} \sigma^{--}_{n_1+n_2-1} + \dots
\nonumber \\
\sigma_{n_1}^{--} \sigma^{++}_{n_2} &=& \frac{n_2}{n_1 (n_1+n_2-1)} \sigma^{++}_{n_1+n_2-1} + \dots
\nonumber \\\sigma_{n_1}^{--} \sigma^{--}_{n_2} &=& \frac{1}{n_1 n_2(n_1+n_2-3)} \sigma^{++}_{n_1+n_2-3} + \dots
\label{LMDoubleOverlap}
\\
\sigma_{n_1}^{--} \sigma^{+-}_{n_2} &=& \frac{1}{n_1} \sigma^{+-}_{n_1+n_2-1} + \dots
\nonumber \\
\sigma_{n_1}^{-+} \sigma^{+-}_{n_2} &=& \frac{1}{n_1+n_2-1} \sigma^{++}_{n_1+n_2-1} + \dots   \label{LMCorrelators}
\end{eqnarray}
We now perform the  renormalization we already saw for the quotient ring 
\begin{equation}
\sigma_{n_i} = n_i \sigma^{--}_{n_i}   
\end{equation}
as well as the renormalizations:
\begin{equation}
\frac{1}{n_i} \sigma_{n_i}^{++} = \text{vol} \, \sigma_{n_i} \, ,
\qquad 
\sigma^{+-}_{n_i} = \alpha^{(2,0)} \sigma_{n_i} \, \qquad
\sigma^{-+}_{n_i} = \alpha^{(0,2)} \sigma_{n_i} \, .
\end{equation}
We note that as long as the volume operator and the unit operator scale oppositely (and the middle cohomology elements do not scale) we keep their two-point functions $T$ invariant.
All the correlation functions (\ref{LMCorrelators}) then precisely match those computed in subsection \ref{CupProductsSingleCycles}. For the special case of the correlator (\ref{LMDoubleOverlap}), this is the case because we concentrate on the term in the co-multiplication that associates the identity operator to the cycle of length one. We note that the correlators computed by elementary means in subsection \ref{CupProductsSingleCycles} already contain a few more results than those that we could extract from the physics results.

\subsection{Cup Products at Low and High Order}
In this section, we compute products of operators at a few low orders $n$. This serves firstly as a small catalogue of  results. Secondly, we will soon see  that these results contain glimpses of  results at large $n$, as well as examples that  inspire. Thus, we 
compute the cup product in the algebra $A \{ S_n \}$ and therefore $A^{[n]}$ for various values of the order $n$ of the permutation group $S_n$. 

\subsubsection{Orders One and Two}
For $n=1$, the cup product equals the ordinary product in the cohomology. This defines the original Frobenius algebra $A$.
For $n=2$, there are two elements in $S_2$. They give rise to the algebras
\begin{eqnarray}
A \{ S_2 \} &=& A^{\otimes 2}  () + A {(12)}
\nonumber \\
A^{[2]} &=& S^2 A [()] + A [(12)]
\end{eqnarray}
with the multiplication rule:
\begin{eqnarray}
\alpha_1 \otimes \alpha_2 () \cdot \beta_1 \otimes \beta_2 () &=& (-1)^{\beta_1 \alpha_2} \alpha_1 \beta_1 \otimes \alpha_2 \beta_2 ()
\nonumber \\
\alpha_1 \otimes \alpha_2 () \cdot \beta (12) &=&  \alpha_1 \alpha_2 \beta (12)
\nonumber \\
\alpha (12) \beta (12) &=& \Delta_\ast (\alpha \beta) ()
\end{eqnarray}
in $A \{ S_2 \}$. An operator $\gamma_1 \otimes \gamma_2$ multiplying the unit element $()$ in the permutation group will be symmetrized in $A^{[2]}$, and otherwise the algebra is similar.

\subsubsection{Order Three}
At order three, we  have the algebras:
\begin{eqnarray}
A \{ S_3 \} &=& A^{\otimes 3}  () + A^{\otimes 2}{(12)} + A (123) + \dots 
\nonumber \\
A^{[2]} &=& S^3 A [()] + A \otimes A [{(12)}] + A [(123)]
\end{eqnarray}
and in the algebra $A \{ S_3 \}$ the multiplication rule is (as in  example 2.17 of \cite{LehnSorgerCupProduct}):
\begin{eqnarray}
(\alpha_1 \otimes \alpha_2) (12).(\beta_1 \otimes \beta_2) (13) &=& \alpha_1 \alpha_2 \beta_1 \beta_2 (132)
\nonumber \\
(\alpha_1 \otimes \alpha_2)(12) . (\beta_1 \otimes \beta_2) (12) &=& (-1)^{\beta_1 \alpha_2 } \Delta_\ast(\alpha_1 \beta_1) \otimes (\alpha_2 \beta_2) ()
\nonumber \\
\alpha (123) . \beta (123) &=& (\alpha \beta e) (132)
\nonumber \\
\alpha(123). \beta(132) &=& \Delta_\ast^{(3)} (\alpha \beta) () \, .
\label{ABracketThree}
\end{eqnarray}
We  used the symbol $ \Delta_\ast^{(3)}$ for the adjoint of the multiplication map $\phi_{3,1}=\Delta^{\ast(3)}: A \otimes A \otimes A \rightarrow A : \alpha_1 \otimes \alpha_2 \otimes \alpha_3 \rightarrow \alpha_1 \alpha_2 \alpha_3 $.
Note that for the first time, we have a multiplication with a graph defect equal to one, leading to a multiplication with the Euler class $e$. Our interpretation of this structure constant as an operator product structure constant makes it the first such three-point function in $AdS_3/CFT_2$ calculated for a covering surface of genus one. We will generalize this example soon.

\subsubsection{Orders Four and Six}
 We have the algebra
\begin{eqnarray}
A^{[4]} &=& A \oplus A \otimes A \oplus S^2 A \oplus S^2 A \otimes  A \oplus S^4 A 
\end{eqnarray}
corresponding  to the  partitions  $[4],[13],[2^2],[1^22],[1^4]$ of $n=4$, and a similar decomposition of the algebra $A^{[6]}$ for $n=6$. We wish to make a few illustrative and useful calculations in these algebras using the multiplications in $A \{ S_{n} \}$. We find the $n=4$ products:
\begin{eqnarray}
\alpha_1 \otimes \alpha_2 \otimes \alpha_3 (12) \cdot 
\beta_1 \otimes \beta_2 \otimes \beta_3 (12)
&=& (-1)^{\beta_1 (\alpha_2 + \alpha_3) + \beta_2 \alpha_3} \Delta_\ast(\alpha_1 \beta_1) \otimes \alpha_2 \beta_2 \otimes \alpha_3 \beta_3  ()
\nonumber 
\\
\alpha_1 \otimes \alpha_2 \otimes \alpha_3 (12) \cdot \beta_1 \otimes \beta_2 \otimes \beta_3 (13) &=& (-1)^{\alpha_3(\beta_1+\beta_2)}
\alpha_1 \alpha_2 \beta_1 \beta_2  \otimes \alpha_3 \beta_3 (132)
\\
\alpha_1 \otimes \alpha_2 \otimes \alpha_3 (12) \cdot \beta_1 \otimes \beta_2 \otimes \beta_3 (34)
&=& (-1)^{(\alpha_2+\alpha_3)(\beta_1+\beta_2)} \alpha_1 \beta_1 \beta_2 \otimes \alpha_2 \alpha_3 \beta_3 (12)(34) \, .
\nonumber 
\end{eqnarray} 
If we compute the analogues in $S_5$, we multiply the fifth wheel on the first wagon with the fifth wheel on the second, but learn little extra. In the algebra $A \{ S_6 \}$ we multiply:
\begin{eqnarray}
\alpha_1 \otimes \dots (123) \beta_1 \otimes \beta_2 \otimes \dots (234) &=& 
\epsilon(\alpha_i,\beta_i) \Delta_\ast(\alpha_1 \alpha_2 \beta_1 \beta_2) \otimes \alpha_3 \beta_3 \otimes \alpha_4 \beta_4 (12)(34)  \nonumber \\
\alpha_1 \otimes \dots (123) \beta_1 \otimes \beta_2 \otimes  \dots  (324) &=& \epsilon \, \Delta_\ast (\alpha_1 \alpha_2 \beta_1 \beta_2) \otimes \dots  (124)
\nonumber 
\\
\alpha_1 \otimes \dots (123) 
\beta_1 \otimes \dots (345) &=& \epsilon \,
\alpha_1 \alpha_2 \alpha_3 \beta_1 \beta_2 \beta_3
\otimes \alpha_4 \beta_4 (12345) \nonumber 
\\
\alpha_1 \otimes \alpha_2 \otimes \dots (123)
\beta_1 \otimes \dots (456) &=&  \epsilon \,
\alpha_1 \beta_1 \beta_2 \beta_3 \otimes \alpha_2 \alpha_3 \alpha_4 \beta_4 (123)(456) 
\, . \label{ABracketSix}
\end{eqnarray}
In these results, we have gradually suppressed  details (like the sign $\epsilon$ depending on the grades of the elements $\alpha_i$ and $\beta_i$ and spectator factors) in order to make them more compact.

The results for the $A \{ S_4 \}$ algebra lead to the gauge invariant $A^{[4]}$ multiplication rule for the transposition conjugaction class:
\begin{eqnarray}
(\alpha_1 \otimes \alpha_2 \otimes \alpha_3 + \alpha_1 \otimes \alpha_3 \otimes \alpha_2) [(12)] 
\cdot (\beta_1 \otimes \beta_2 \otimes \beta_3 + \beta_1 \otimes \beta_3 \otimes \beta_2)[(12)] &=&  \nonumber \\
 6 \Big( (-1)^{\beta_1 (\alpha_2 + \alpha_3) + \beta_2 \alpha_3} \Delta_\ast(\alpha_1 \beta_1) \otimes \alpha_2 \beta_2 \otimes \alpha_3 \beta_3 + \text{three  symm.} \Big) [()] & &
\nonumber \\
+ 2 \Big( (-1)^{(\alpha_2+\alpha_3)(\beta_1+\beta_2)} \alpha_1 \beta_1 \beta_2 \otimes \alpha_2 \alpha_3 \beta_3
+ \text{three symm.}  \Big)
[(12)][(34)]  &&
\nonumber \\
+ 3 \Big( \alpha_1 \alpha_2 \beta_1 \beta_2 + \text{three symmetrization terms}  \Big) [(123)]  \, . &&
\end{eqnarray}
It is clear that the structure constants depend on the seed theory only, though the combinatorics depends on the order of the symmetric group. 
In the algebra of gauge invariants $A^{[6]}$ we recall that the $[1^3 3]$ partition corresponds to a term $S^3 A \otimes A$ in the direct sum $A^{[6]}$. When we compute the product of three-cycle conjugacy classes, we therefore symmetrize the results (\ref{ABracketSix}) as well as (\ref{ABracketThree}) over the factors associated to the cycles of length one, giving rise to six terms for each element of the conjugacy class $[1^3 3]$, of which there are  $40$. Thus, a naive counting gives $240^2$ terms. While this can be written significantly more efficiently, we still refrain from writing out the final result. We remark that the conjugacy class combinatorics is captured by the product:
\begin{equation}
{[} 1^3 3] \ast [1^3 3] = 40 [1^6] + 8 [1^2 2^2] + 10 [1^3 3] + 2 [3^2] +5 [15] \, .
 \end{equation}

\subsubsection{Remarks}
\begin{itemize}
\item
We note that when we express the ring structure in an 
appropriate set of cohomology elements, the structure constants  become independent of the order of the orbifold $n$ \cite{LQWStability}, such that even low order $n$ results capture structure constants of the large order ring (as long as the order is large enough to accommodate all possible conjugacy classes that can occur in the product). We  refer to \cite{LQWStability} for the precise statement.  This property  holds  in the presence of genus one Euler class corrections. 
\item  The combinatorial description of top connection coefficients and sub-top coefficients of the symmetric group \cite{GoupilBedard,GouldenJacksonCombinatorial} may be useful to efficiently compute structure constants of the Hilbert scheme of points for compact surfaces.
\end{itemize}

\subsection{Genus One and an Overlap of Three}
In this  subsection, we compute a simple yet new class of structure constants. We consider two operators $\alpha \pi$ and $\beta \rho$, proportional to permutations 
$\pi=\sigma_{(n_1)}=(1234\dots n_1)$ and $\rho=\sigma_{(n_2)}=
(1,2,3 ,n_1+1 ,n_1+2 ,\dots ,n_1+n_2-3)$ which overlap in the three elements $1,2,3$. We list the orbits:
\begin{eqnarray}
\langle \pi \rangle \setminus [n_1+n_2-3] &=& \{ [n_1],n_1+1,n_1+2,\dots, n_1+n_2-3 \}
\nonumber \\
\langle \rho \rangle \setminus [n_1+n_2-3] &=& \{ 4,5,\dots,n_1, \{1,2,3,n_1+1,n_1+2,\dots,n_1+n_2-3 \} \}
\nonumber \\
\langle \pi \rho \rangle \setminus [n_1+n_2-3]&=& \{1, 2, \dots, n_1+n_2-3  \}
\nonumber \\
\langle \pi , \rho \rangle \setminus [n_1+n_2-3] &=& [n_1+n_2-3] \, .
\end{eqnarray}
Importantly, the graph defect equals
\begin{equation}
g = \frac{1}{2} ( n_1+n_2-1 -(n_2-2)-(n_1-2)-1)=1 \, .
\end{equation}
We find the operator product:
\begin{equation}
\alpha \pi \cdot \beta \rho = (\alpha \beta e) \pi \rho \, .
\end{equation}
This is a  generalization of the product we found for equal three-cycles in $S_3$. It is a simple and generic genus one result for the operator product expansion in the topological orbifold conformal field theory that forms a  clear target for the holographically dual bulk $AdS_3$ topological string theory.

\subsection{Further Examples}

It is straightforward to generate further classes of examples using the tools we have laid out. 
\begin{itemize}
\item Note that for  $2l+1$ identical (consecutive) elements in two cyclic permutations, the graph defect equals $g=l$, and therefore the product vanishes for $l=g \ge 2$. This provides an example of how to generate (vanishing) higher genus amplitudes of arbitrarily high genus.
\item An example that involves the co-multiplication map $\Delta_\ast^{(n)}$, namely the adjoint of $\phi_{n,1}$, is found in the multiplication of two elements associated to inverse $n$-cycles: \\
$\alpha (12 \dots n) \cdot \beta (n,n-1,\dots,1) = \Delta_\ast^{(n)} (\alpha \beta) () \, .$
 \item Co-multiplication can also act on a graph defect one contribution:\\
 $ \alpha (123 \dots n) \cdot \beta (1,2,n,n-1, \dots 4,3)
 = \Delta_\ast^{(n-2)} (\alpha \beta e) (132) \, .$
\item Finally, we provide an example involving a permutation that is the product of two cycles in the algebra $A \{ S_4 \}$: \\
$ \alpha_1 \otimes \alpha_2 (12)(34) \cdot \beta_1 \otimes \beta_2 \otimes \beta_3 (23) = \alpha_1 \alpha_2 \beta_1 \beta_2 \beta_3 (1243) \, .$
\end{itemize}
We leave room to further combine and invent examples in creative ways.

\section{Conclusions}
\label{Conclusions}

We described the topological symmetric orbifold conformal field theory on a complex surface $M$ on the basis of the mathematics literature on the Hilbert scheme of points on the surface. We started out by studying a universal quotient of the operator ring, which is isomorphic to the cohomology ring of the Hilbert scheme of points on the complex two-plane $\mathbb{C}^2$. The mathematical description of the ring \cite{LehnSorgerSymmetric} allowed to   leverage the literature on permutation group combinatorics to provide explicit expressions for correlators. Moreover, we were able to  prove a conjectured description of  a set of extremal correlators  in terms of Hurwitz numbers \cite{Pakman:2009ab}. 

The full cohomology ring is  compactly captured by a symmetric Frobenius algebra described in \cite{LehnSorgerCupProduct}. We showed that the correlators that one computes in the symmetric orbifold conformal field theory as well as the bulk string theory agree with the mathematical description of the orbifold cohomology ring. The calculations are very efficient in the mathematical formalism, which uses only the structure constants of the seed cohomology ring and combinatorics. We stressed that the structure constants in the topologically twisted theory can be rendered independent of the order of the orbifold as long as one picks it to be (finite but) large enough. We moreover computed a set of genus one correlators and observed that connected higher genus correlators vanish.

The simplification of the topological symmetric orbifold theory is a good preparation for proving a topological AdS/CFT correspondence. The task is to bijectively map the bulk gauge invariants to the operator algebra $A^{[n]}$ that captures the boundary chiral ring. For the quotient ring, the problem reduces to identifying bulk operators that map bijectively into the conjugacy classes of the symmetric group $S_n$ (while ignoring all non-trivial topology in the manifold $M$). Once the correct bijection has been identified, the matching of the structure constants may well be manageable. To establish the bijection, one takes cues from the matching of spectra generically (see e.g. \cite{Argurio:2000tb}) or from the example at level $k=1$ \cite{Eberhardt:2019ywk,Eberhardt:2020akk}, namely when the bulk curvature radius equals the string length.

Moreover, one would like to directly twist the bulk string theory, and gather further insight into the topological AdS/CFT duality. This can be done following one of the approaches of \cite{Sugawara:1999fq,Rastelli:2005ph,Li:2019qzx,Costello:2020jbh,Li:2020nei}  and should lead to an alternative and interesting representation of the Frobenius algebra at hand. The topological bulk string theory correlators should give rise, after localization, to an intersection theory on the moduli space of Riemann surfaces that must match the space-time cohomology ring (and in its simplest incarnation, the Hurwitz numbers). Holomorphic covering maps are bound to play an important role \cite{Rastelli:2005ph,Pakman:2009zz,Eberhardt:2019ywk,Eberhardt:2020akk}.

Importantly, we can propose a concrete mathematical incarnation of the proof of holographic duality (for simplicity in the case of $M=\mathbb{C}^2$ though the relation holds more generally). An equivalence between the Gromov-Witten theory on $\mathbb{P}^1 \times \mathbb{C}^2$ and
the quantum cohomology of the Hilbert scheme of points on $\mathbb{C}^2$ was proven in \cite{OP}. We propose that this 
is a precise mathematical realizaton of a non-perturbative proof of a topological AdS/CFT duality. The quantum cohomology ring of the Hilbert scheme is the ring of the topologically twisted boundary theory, including  instanton corrections which are responsible for the quantum deformation. The closed topological A-model on the space $\mathbb{P}^1 \times \mathbb{C}^2$ is to be identified with the topologically twisted bulk dual. This proposal is compelling, and the equivalence theorem proven in \cite{OP} provides an explicit holographic map.

Finally, a duality symmetry in string theory, uncovered in \cite{Dijkgraaf:1998gf} and further analyzed in \cite{deBoer:2008ss,Bourget:2015ffp}, predicts a covariance of the ring of the topological symmetric orbifold theory on $T^4$ or $K3$ that goes beyond what is known about the structure constants in the mathematics literature. It would be interesting to prove the duality covariance of the ring. 

\section*{Acknowledgments}
It is a pleasure to thank our colleagues for creating a stimulating research environment.

\appendix

\section{Operator Product Coefficients}
\label{LuninMathurExtraction}
\label{LuninMathurCorrelators}
\label{OperatorProductCoefficients}
In this appendix only, for easier comparison, we use the notation of \cite{Lunin:2001pw}, and we extract from that paper a number of structure constants that we use in the bulk of the paper.
The paper defines and uses operators $\sigma^{\pm,\pm}_{n}$ of (left, right) R-charge $n \pm 1$ depending on the sign of the upper index on the operator.
The structure constants in the operator product expansion of two of these operators giving rise to a third are computed separately for right- and left-movers. Moreover, the paper \cite{Lunin:2001pw} separates out a reduced structure constant where one has factored out an $su(2)_R$ Clebsch-Gordan coefficient. This is sensible if one wishes to capture the full $N=(4,4)$ superconformal structure of the correlators. In twisting the topological theory, we pick a particular $N=(2,2)$ superconformal subalgebra, and we break the $su(2)_R$ symmetry. We therefore wish to write the structure constants rather in a $u(1)_R$ language (and restore the Clebsch-Gordan coefficients). 
Firstly, we recall from \cite{Lunin:2001pw} the $su(2)_R$ reduced chiral structure constants:
\begin{eqnarray}
\hat{C}^{1_n,1_m,1_q}_{nmq} &=& \left(
\frac{(1_n n+ 1_m m +1_q q +1)^2}{4 m n q} \frac{\Sigma! \alpha_n! \alpha_m! \alpha_q!}{(n+1_n)! (m+1_m)! (q+1_q)!}\right)^{\frac{1}{2}}
\nonumber \\
\Sigma &=& \frac{1}{2} (n+1_n + m + 1_m + q + 1_q) +1 
\nonumber \\
\alpha_n &=& \Sigma-n-1_n-1 \, .
\end{eqnarray}
If we delve into appendix C of \cite{Lunin:2001pw}, we can  reconstruct the $N=2$ structure constants from these $N=4$ structure constants by multiplying back in the Clebsch-Gordan coefficient:
\begin{eqnarray}
\left( \begin{array}{ccc}
\frac{n+1_n}{2} & \frac{m+1_m}{2} & \frac{q+1_q}{2} \\
\frac{q+1_q-m-1_m}{2} & \frac{m+1_m}{2} & - \frac{q+1_q}{2} 
\end{array} \right)
&=& \left( \frac{(m+1_m)! (q+1_q)!}{
\Sigma! \alpha_n!
} \right)^{\frac{1}{2}}
\end{eqnarray}
We then obtain the formula:
\begin{eqnarray}
{C}^{1_n,1_m,1_q}_{nmq} &=& \left(
\frac{(1_n n+ 1_m m +1_q q +1)^2}{4 m n q} \frac{ \alpha_m! \alpha_q!}{(n+1_n)! }\right)^{\frac{1}{2}}
 \\
\Sigma &=& \frac{1}{2} (n+1_n + m + 1_m + q + 1_q) +1 
\nonumber \\
\alpha_n &=& \Sigma-n-1_n-1 \, .
\end{eqnarray}
Since we have the $U(1)_R$ charge constraint $n+1_n+m+1_m=q+1_q$, we can simplify further and obtain the equation 
(\ref{NonReducedStructureConstant}) for the chiral structure constants in the bulk of the paper.

\bibliographystyle{JHEP}

\begin{thebibliography}{99}

\bibitem{Maldacena:1997re}
J.~M.~Maldacena,
``The Large N limit of superconformal field theories and supergravity,''
Int. J. Theor. Phys. \textbf{38} (1999), 1113-1133
doi:10.1023/A:1026654312961
[arXiv:hep-th/9711200 [hep-th]].

\bibitem{Dabholkar:2007ey}
A.~Dabholkar and A.~Pakman,
``Exact chiral ring of AdS(3) / CFT(2),''
Adv. Theor. Math. Phys. \textbf{13} (2009) no.2, 409-462
doi:10.4310/ATMP.2009.v13.n2.a2
[arXiv:hep-th/0703022 [hep-th]].

\bibitem{Gaberdiel:2007vu}
M.~R.~Gaberdiel and I.~Kirsch,
``Worldsheet correlators in AdS(3)/CFT(2),''
JHEP \textbf{04} (2007), 050
doi:10.1088/1126-6708/2007/04/050
[arXiv:hep-th/0703001 [hep-th]].


\bibitem{Eberhardt:2019ywk}
  L.~Eberhardt, M.~R.~Gaberdiel and R.~Gopakumar,
  ``Deriving the $\text{AdS}_{3}/\text{CFT}_{2}$ Correspondence,''
  JHEP {\bf 2002} (2020) 136
  doi:10.1007/JHEP02(2020)136
  [arXiv:1911.00378 [hep-th]].

\bibitem{Eberhardt:2020akk}
L.~Eberhardt,
``AdS$_{3}$/CFT$_{2}$ at higher genus,''
JHEP \textbf{05} (2020), 150
doi:10.1007/JHEP05(2020)150
[arXiv:2002.11729 [hep-th]].

\bibitem{Jevicki:1998bm}
A.~Jevicki, M.~Mihailescu and S.~Ramgoolam,
``Gravity from CFT on S**N(X): Symmetries and interactions,''
Nucl. Phys. B \textbf{577} (2000), 47-72
doi:10.1016/S0550-3213(00)00147-4
[arXiv:hep-th/9907144 [hep-th]].


\bibitem{Lunin:2000yv}
O.~Lunin and S.~D.~Mathur,
``Correlation functions for M**N / S(N) orbifolds,''
Commun. Math. Phys. \textbf{219} (2001), 399-442
doi:10.1007/s002200100431
[arXiv:hep-th/0006196 [hep-th]].

  
\bibitem{Lunin:2001pw}
O.~Lunin and S.~D.~Mathur,
``Three point functions for M(N) / S(N) orbifolds with N=4 supersymmetry,''
Commun. Math. Phys. \textbf{227} (2002), 385-419
doi:10.1007/s002200200638
[arXiv:hep-th/0103169 [hep-th]].

\bibitem{Witten:1988ze}
  E.~Witten,
  ``Topological Quantum Field Theory,''
  Commun.\ Math.\ Phys.\  {\bf 117} (1988) 353.
  doi:10.1007/BF01223371
 
 
  
  
  
  
  
  
\bibitem{Witten:1988xj}
  E.~Witten,
  ``Topological Sigma Models,''
  Commun.\ Math.\ Phys.\  {\bf 118} (1988) 411.
  doi:10.1007/BF01466725
  
\bibitem{Eguchi:1990vz}
  T.~Eguchi and S.~K.~Yang,
  ``N=2 superconformal models as topological field theories,''
  Mod.\ Phys.\ Lett.\ A {\bf 5} (1990) 1693.
  doi:10.1142/S0217732390001943
  
  
\bibitem{Rastelli:2005ph}
  L.~Rastelli and M.~Wijnholt,
  ``Minimal AdS(3),''
  Adv.\ Theor.\ Math.\ Phys.\  {\bf 11} (2007) no.2,  291
  doi:10.4310/ATMP.2007.v11.n2.a4
  [hep-th/0507037].
 
  
\bibitem{Pakman:2009ab}
A.~Pakman, L.~Rastelli and S.~S.~Razamat,
``Extremal Correlators and Hurwitz Numbers in Symmetric Product Orbifolds,''
Phys. Rev. D \textbf{80} (2009), 086009
doi:10.1103/PhysRevD.80.086009
[arXiv:0905.3451 [hep-th]].

  

\bibitem{Sugawara:1999fq}
Y.~Sugawara,
``Topological string on AdS(3) x N,''
Nucl. Phys. B \textbf{576} (2000), 265-284
doi:10.1016/S0550-3213(00)00075-4
[arXiv:hep-th/9909146 [hep-th]].

  
  
  
  
\bibitem{Li:2019qzx}
S.~Li and J.~Troost,
``Pure and Twisted Holography,''
JHEP \textbf{03} (2020), 144
doi:10.1007/JHEP03(2020)144
[arXiv:1911.06019 [hep-th]].


\bibitem{Costello:2020jbh}
K.~Costello and N.~M.~Paquette,
``Twisted Supergravity and Koszul Duality: A case study in AdS$_3$,''
[arXiv:2001.02177 [hep-th]].


\bibitem{Li:2020nei}
S.~Li and J.~Troost,
``Twisted String Theory in Anti-de Sitter Space,''
[arXiv:2005.13817 [hep-th]].


\bibitem{Grothendieck}
A.~Grothendieck, ``Fondements de la G\'eom\'etrie Alg\'ebrique, S\'eminaire Bourbaki," IHP, Paris
(1957-1964)
\bibitem{Fogarty}
J.~Fogarty, ``Algebraic families on an Algebraic Surface," Amer. Jour. of Math. {bf 13}, No 2, (1968),
511–521


\bibitem{Beauville}
A.~Beauville, ``Variétés Kähleriennes dont la premiere classe de Chern est nulle," Journal of Differential Geometry, {\bf 18}(4), 755-782 (1983).


\bibitem{Briancon}
J.~Briancon, ``Description of $Hilb^n \mathbb{C}[x,y]$, Invent. Math. {\bf 41} (1977), 45–89

\bibitem{EllingsrudStroemme}
G.~Ellingsrud,  S.~A.~Strømme,  ``On the homology of the Hilbert scheme of points in the plane," Inventiones mathematicae, {\bf 87}(2), 343-352 (1987).

\bibitem{NakajimaLectures}
H.~Nakajima, ``Lectures on Hilbert schemes of points on surfaces (No. 18). American Mathematical Soc. (1999).


\bibitem{Grojnowski}
I.~Grojnowski,  ``Instantons and affine algebras I: the Hilbert scheme and vertex operators," arXiv preprint alg-geom/9506020.

\bibitem{Lehn}
M.~Lehn, ``Chern classes of tautological sheaves on Hilbert schemes of points on surfaces," Inventiones mathematicae, {\bf 136}(1), 157-207  (1999). 

\bibitem{LehnSorgerSymmetric}
M.~Lehn, C.~Sorger, ``Symmetric groups and the cup product on the cohomology of Hilbert schemes," Duke Mathematical Journal {\bf 110.2} (2001): 345-357.

\bibitem{Vasserot}
E.~Vasserot, ``Sur l'anneau de cohomologie du schéma de Hilbert de C2," Comptes Rendus de l'Académie des Sciences-Series I-Mathematics, {\bf 332}(1), 7-12  (2001). 





\bibitem{Goettsche}
L.~Göttsche, ``The Betti numbers of the Hilbert scheme of points on a smooth projective surface," Mathematische Annalen, {\bf 286}(1-3), 193-207 (1990). 

\bibitem{LehnSorgerCupProduct}
M.~Lehn, C.~Sorger, ``The cup product of Hilbert schemes for K3 surfaces," Inventiones mathematicae {\bf 152.2} (2003): 305-329.


\bibitem{FantechiGoettscheOrbifoldCohomology}
B.~Fantechi, L.~Göttsche, ``Orbifold cohomology for global quotients," Duke Mathematical Journal {\bf 117.2} (2003): 197-227.


\bibitem{Chen:2000cy}
W.~m.~Chen and Y.~b.~Ruan,
``A New cohomology theory for orbifold,''
Commun. Math. Phys. \textbf{248} (2004), 1-31
doi:10.1007/s00220-004-1089-4
[arXiv:math/0004129 [math.AG]].

\bibitem{Ruan:2000mz}
Y.~Ruan,
``Stringy geometry and topology of orbifolds,''
[arXiv:math/0011149 [math.AG]].




\bibitem{Bertin}
J.~Bertin, ``The punctual Hilbert scheme: an introduction,"
 (2008).






\bibitem{Dixon:1985jw}
L.~J.~Dixon, J.~A.~Harvey, C.~Vafa and E.~Witten,
``Strings on Orbifolds,''
Nucl. Phys. B \textbf{261} (1985), 678-686
doi:10.1016/0550-3213(85)90593-0

\bibitem{Dixon:1986jc}
L.~J.~Dixon, J.~A.~Harvey, C.~Vafa and E.~Witten,
``Strings on Orbifolds. 2.,''
Nucl. Phys. B \textbf{274} (1986), 285-314
doi:10.1016/0550-3213(86)90287-7


\bibitem{Dijkgraaf:1996xw}
R.~Dijkgraaf, G.~W.~Moore, E.~P.~Verlinde and H.~L.~Verlinde,
``Elliptic genera of symmetric products and second quantized strings,''
Commun. Math. Phys. \textbf{185} (1997), 197-209
doi:10.1007/s002200050087
[arXiv:hep-th/9608096 [hep-th]].


\bibitem{Haiman98}
M.~Haiman,``t, q-Catalan numbers and the Hilbert scheme," Discrete Mathematics, {\bf 193} (1-3), 201-224 (1998).

\bibitem{Polchinski2}
J~Polchinski, ``String theory: Volume 2, superstring theory and beyond," Cambridge university press (1998).



\bibitem{JamesKerber}
G.~James, A.~Kerber, ``The Representation Theory of the Symmetric Group,"  Encyclopedia Math. Appl. (1981).

\bibitem{GAP}
The GAP Group, GAP -- Groups, Algorithms, and Programming, Version 4.11.0; 2020. (https://www.gap-system.org)


\bibitem{IvanovKerov}
V.~Ivanov, S.~Kerov,  ``The algebra of conjugacy classes in symmetric groups and partial permutations," Journal of Mathematical Sciences, {\bf 107} (5), 4212-4230  (2001).



\bibitem{Goulden}
I.~Goulden, ``A differential operator for symmetric functions and the combinatorics of multiplying transpositions," Transactions of the American Mathematical Society, {\bf 344}(1), 421-440  (1994).


\bibitem{Goupil}
A.~Goupil, ``On products of conjugacy classes of the symmetric group," Discrete mathematics, {\bf 79} (1), 49-57 (1990).

\bibitem{FarahatHigman}
H.~Farahat, G.~Higman, ``The centres of symmetric group rings," Proceedings of the Royal Society of London. Series A. Mathematical and Physical Sciences, {\bf 250} (1261), 212-221  (1959). 

\bibitem{GoupilBedard}
F.~Bédard,  A.~Goupil, ``The poset of conjugacy classes and decomposition of products in the symmetric group," Canadian mathematical bulletin, {\bf 35} (2), 152-160(1992).
\bibitem{GouldenJacksonCombinatorial}
I.~Goulden, D.~Jackson, ``The combinatorial relationship between trees, cacti and certain connection coefficients for the symmetric group," European Journal of Combinatorics, {\bf 13}(5), 357-365 (1992).


\bibitem{GouldenJacksonSymmetric}
I.~Goulden, D.~Jackson,``Symmetrical Functions and Macdonald′s Result for Top Connexion Coefficients in the Symmetrical Group," Journal of Algebra, {\bf 166} (2), 364-378  (1994). 





\bibitem{Dijkgraaf:2003nw}
R.~Dijkgraaf and L.~Motl,
``Matrix string theory, contact terms, and superstring field theory,''
[arXiv:hep-th/0309238 [hep-th]].


  \bibitem{LQW2005}
 W.~P.~Li, Z.~Qin, W.~Wang, ``Hilbert scheme intersection numbers, Hurwitz numbers, and Gromow-Witten invariants", In Infinite-Dimensional Aspects of Representation Theory and Applications: International Conference on Infinite-Dimensional Aspects of Representation Theory and Applications, May 18-22, 2004, University of Virginia, Charlottesville, Virginia (Vol. 392, p. 67). American Mathematical Soc.
  
  
\bibitem{Hurwitz}
A.~Hurwitz, ``Über Riemann'sche Flächen mit gegebenen Verzweigungspunkten," Mathematische Annalen, {\bf 39} (1), 1-60  (1891).
  
  \bibitem{LQW2004}
 W.~P.~Li,  Z.~Qin, W.~Wang, ``Ideals of the cohomology rings of Hilbert schemes and their applications," Transactions of the American Mathematical Society, {\bf 356} (1), 245-265.

\bibitem{Pakman:2009zz}
A.~Pakman, L.~Rastelli and S.~S.~Razamat,
``Diagrams for Symmetric Product Orbifolds,''
JHEP \textbf{10} (2009), 034
doi:10.1088/1126-6708/2009/10/034
[arXiv:0905.3448 [hep-th]].


 
 

\bibitem{LQWStability}
W.~P.~Li,  Z.~Qin, W.~Wang, ``Stability of the cohomology rings of Hilbert schemes of points on surfaces," arXiv preprint math/0107139.

\bibitem{Argurio:2000tb}
R.~Argurio, A.~Giveon and A.~Shomer,
``Superstrings on AdS(3) and symmetric products,''
JHEP \textbf{12} (2000), 003
doi:10.1088/1126-6708/2000/12/003
[arXiv:hep-th/0009242 [hep-th]].

\bibitem{OP}
A.~Okounkov, R.~!Pandharipande, ``Quantum cohomology of the Hilbert scheme of points in the plane," Inventiones mathematicae, \textbf{179} (3)  (2010), 523-557.


\bibitem{Dijkgraaf:1998gf}
R.~Dijkgraaf,
``Instanton strings and hyperKahler geometry,''
Nucl. Phys. B \textbf{543} (1999), 545-571
doi:10.1016/S0550-3213(98)00869-4
[arXiv:hep-th/9810210 [hep-th]].

\bibitem{deBoer:2008ss}
J.~de Boer, J.~Manschot, K.~Papadodimas and E.~Verlinde,
``The Chiral ring of AdS(3)/CFT(2) and the attractor mechanism,''
JHEP \textbf{03} (2009), 030
doi:10.1088/1126-6708/2009/03/030
[arXiv:0809.0507 [hep-th]].

\bibitem{Bourget:2015ffp}
A.~Bourget and J.~Troost,
``The Covariant Chiral Ring,''
JHEP \textbf{03} (2016), 163
doi:10.1007/JHEP03(2016)163
[arXiv:1512.03649 [hep-th]].


\end{thebibliography}

\end{document}